\documentclass[english,aps,prb,amssymb,amsbsy,amsmath,showpacs,superscriptaddress,preprint]{revtex4}
\usepackage[T1]{fontenc}
\usepackage[latin9]{inputenc}
\usepackage{textcomp}
\usepackage{bm}
\usepackage{amsmath}
\usepackage{amssymb}
\usepackage{graphicx}
\usepackage{esint}

\makeatletter

\providecommand{\tabularnewline}{\\}

\@ifundefined{textcolor}{}
{%
 \definecolor{BLACK}{gray}{0}
 \definecolor{WHITE}{gray}{1}
 \definecolor{RED}{rgb}{1,0,0}
 \definecolor{GREEN}{rgb}{0,1,0}
 \definecolor{BLUE}{rgb}{0,0,1}
 \definecolor{CYAN}{cmyk}{1,0,0,0}
 \definecolor{MAGENTA}{cmyk}{0,1,0,0}
 \definecolor{YELLOW}{cmyk}{0,0,1,0}
 }

\makeatother

\usepackage{babel}
\begin{document}

\title{First-principles calculation of topological invariants $Z_{2}$ within
the FP-LAPW formalism}

\author{Wanxiang Feng}

\affiliation{Beijing National Laboratory for Condensed Matter Physics and Institute
of Physics, Chinese Academy of Sciences, Beijing 100190, China}

\affiliation{Materials Science \& Technology Division, Oak Ridge National Laboratory,
Oak Ridge, Tennessee 37831, USA}

\affiliation{Department of Physics and Astronomy, University of Tennessee, Knoxville,
Tennessee 37996, USA}

\author{Jun Wen}

\affiliation{Department of Physics, University of Texas at Austin, Austin, Texas
78712, USA}

\author{Jinjian Zhou}

\affiliation{Beijing National Laboratory for Condensed Matter Physics and Institute
of Physics, Chinese Academy of Sciences, Beijing 100190, China}

\author{Di Xiao}

\affiliation{Materials Science \& Technology Division, Oak Ridge National Laboratory,
Oak Ridge, Tennessee 37831, USA}

\author{Yugui Yao}

\email{ygyao@iphy.ac.cn}

\selectlanguage{english}%

\affiliation{Beijing National Laboratory for Condensed Matter Physics and Institute
of Physics, Chinese Academy of Sciences, Beijing 100190, China}

\date{\today}
\begin{abstract}
In this paper, we report the implementation of first-principles calculations
of topological invariants $Z_{2}$ within the full-potential linearized
augmented plane-wave (FP-LAPW) formalism. In systems with both time-reversal
and spatial inversion symmetry (centrosymmetric), one can use the
parity analysis of Bloch functions at time-reversal invariant momenta
to determine the $Z_{2}$ invariants. In systems without spatial inversion
symmetry (noncentrosymmetric), however, a more complex and systematic
method in terms of the Berry gauge potential and the Berry curvature
is required to identify the band topology. We show in detail how both
methods are implemented in FP-LAPW formalism and applied to several
classes of materials including centrosymmetric compounds Bi$_{2}$Se$_{3}$
and Sb$_{2}$Se$_{3}$ and noncentrosymmetric compounds LuPtBi, AuTlS$_{2}$
and CdSnAs$_{2}$. Our work provides an accurate and effective implementation
of first-principles calculations to speed up the search of new topological
insulators.
\end{abstract}


\maketitle

\section{introduction}

Recently, topological insulators (TIs) have attracted great attention
in the fields of condensed matter physics and materials science. Based
on the noninteracting band theory, TIs have gapped bulk gap and time-reversal
symmetry protected metallic helical surface (edge) states where spin
and momentum are locked together.\cite{Kane2005a,Kane2005b} These
novel physical properties hold great promise in applications of spintronics
and quantum computing\cite{Moore2010} and have stimulated both experimental
and theoretical studies. Indeed, the field of TIs is expanding so
rapidly and there have been several excellent review articles on it.\cite{Qi2010,Qi2011,Hasan2010}
Although many TIs including alloy\cite{Teo2008,Hsieh2008}, binary
compounds\cite{Bernevig2006,Konig2007,Fu2007a,Zhang2009,Xia2009,Chen2009},
ternary compounds\cite{Xiao2010a,Chadov2010,Lin2010a,Lin2010b,Yan2010a,Chen2010,Sato2010,Kuroda2010,Sun2010,Yan2010b,Kim2010,Feng2011,Jin2011,Zhang2011},
and quaternary compounds\cite{Chen2011} have already been theoretically
predicted and experimentally realized, real materials that can be
used in practical engineering are still needed. Therefore searching
for new TIs with a variety of excellent physical properties has become
a central task in this filed. To achieve this goal, one has to develop
an accurate and effective method to distinguish TIs from normal insulators.

There are several general methods to determine the band topology of
an insulator:

(i) Based on the idea of bulk-edge correspondence of TIs,\cite{Moore2010,Qi2010,Qi2011,Hasan2010}
one can calculate surface (edge) states for a given insulator and
count the number of gapless modes across the Fermi level. An odd number
of gapless modes implies a TI while an even number indicates a normal
insulator. This is a straightforward but not efficient way because
the surface state dispersion may depend on every detail of the surface,
for example, grown directions, terminated chemical elements and surface
reconstructions. In some materials, the topologically nontrivial and
trivial surface states can coexist, which further complicates the
identification of the bulk topological order. To make sure that the
gapless modes are topologically protected, one has to vary surface
crystal structures and see if gapless modes can survive. Furthermore,
a huge amount of computational resources is required in first-principles
surface calculations.

(ii) It is possible to use adiabatic continuity and so-called band
inversion mechanism to identify TIs.\cite{Chadov2010,Lin2010a,Lin2010b,Yan2010a,Feng2011,Liu2011}
The adiabatic continuity can be realized by artificially changing
some external parameter such as the spin-orbit coupling (SOC) strength
or lattice constant. Suppose the unknown state is near some known
topological trivial or nontrivial states in a parameter space. If
one tunes the parameter and the band gap stays open until it reaches
the known state, then by the principle of adiabatic continuity, these
two states share the same topological classification. Otherwise, the
unknown state and known state may have different topological classifications
if the band gap closes. Obviously, many intermediate calculations
are required, making it a very tedious work. Band inversion at high-symmetry
points within the Brillouin zone (BZ), as an empirical rule, can also
be used to reveal the band topology. Although this empirical rule
is adapted in some materials, such as half-Heusler\cite{Xiao2010a,Chadov2010,Lin2010a},
chalcogenide\cite{Lin2010b,Yan2010a}, and chalcopyrite\cite{Feng2011}
compounds, it is not an universal way for arbitrary systems.

(iii) The most general and direct approach is to calculate $Z_{2}$
topological invariants from the knowledge of Bloch band theory.\cite{Fu2007b,Moore2007,Roy2009}
For materials with both time-reversal and spatial inversion symmetry
(centrosymmetric systems), the simple parity criterion developed by
Fu and Kane\cite{Fu2007a} have been applied in a number of works.\cite{Zhang2009,Yan2010a,Yan2010b,Kim2010,Zhang2011}
On the other hand, if the spatial inversion symmetry is absent (noncentrosymmetric
systems), one must resort to a more complex method to evaluate $Z_{2}$
invariants.\cite{Fu2006} Within a tight-binding framework, Fukui
and Hatsugai\cite{Fukui2007} have developed an effective algorithm
to compute $Z_{2}$ invariants in terms of the Berry gauge potential
and the Berry curvature\cite{Xiao2010b} associated with the Bloch
functions (BFs). This method has already been implemented in our first-principles
codes and successfully predicted three-dimensional (3D) TIs in ternary
half-Heusler\cite{Xiao2010a} and chalcopyrite\cite{Feng2011} compounds
and two-dimensional (2D) quantum spin hall effect (QSHE) in Silicene.\cite{Liu2011}
Recently there appears another method which is in the same spirit
of Ref. \onlinecite{Fu2006} but employs the charge center of Wannier
functions.\cite{Soluyanov2011,Yu2011}

In this work, we illustrate the detailed implementation of first-principles
calculations of topological invariants $Z_{2}$ in both centrosymmetric
and noncentrosymmetric systems within the full-potential linearized
augmented plane-wave (FP-LAPW) formalism. Although the latter method
for noncentrosymmetric systems can be applied to centrosymmetric systems,
the parity criterion for centrosymmetric systems is a simpler and
quicker way to determine the band topology. For this reason, we here
introduce both of these methods. It should be emphasized that our
methods are standard post-process after ground state wavefunctions
are obtained in self-consistent calculation, so the calculation of
$Z_{2}$ invariants becomes a routine task just like band structures
and density of states. Additionally, we have already paralleled our
first-principles codes to speed up the calculation. Our implementation
of the calculation of $Z_{2}$ invariants is expected to be an efficient
way for searching new TIs.

The paper is organized as follows. In Sec. II, we review the fundamental
expression of BFs within FP-LAPW formalism and the construction of
overlap matrix, and then give the detailed formalism for implementation
of parity analysis in centrosymmetric systems and lattice calculation
of $Z_{2}$ invariants in noncentrosymmetric systems. In Sec. III,
we take centrosymmetric compounds Bi$_{2}$Se$_{3}$ and Sb$_{2}$Se$_{3}$
and noncentrosymmetric compounds LuPtBi, AuTlS$_{2}$ and CdSnAs$_{2}$
to illustrate the efficiency of our methods. Finally, we give a brief
summary of our work in Sec. IV. In App. A, we provide details on the
overlap matrix and its derivatives.

\section{methods}

In this section, we start by reviewing the formalism of BFs within
FP-LAPW formalism and the construction of overlap matrix,\cite{Singh1994,Blaha2001,Blugel2006}
then illustrate the calculation of $Z_{2}$ invariants in both centrosymmetric
and noncentrosymmetric systems. The key is to calculate the eigenvalues
of parity operator according to parity criterion \cite{Fu2007a} (in
the former case) or the overlap matrices related to time-reversal
operator \cite{Fukui2007} (in the latter case).

\subsection{Bloch functions and overlap matrix\label{sub:II A}}

In the case of SOC, we consider BFs with two components,

\begin{equation}
\Psi_{n\bm{k}}\left(\bm{r}\right)=\left[\begin{array}{c}
\psi_{n\bm{k}}^{\uparrow}\left(\bm{r}\right)\\
\psi_{n\bm{k}}^{\downarrow}\left(\bm{r}\right)
\end{array}\right],
\end{equation}
 where $\uparrow$ and $\downarrow$ refer to the up and down component
of spin. The periodic part of BFs is $u_{n\bm{k}}\left(\bm{r}\right)=e^{-i\bm{k}\cdot\bm{r}}\left[\begin{array}{cc}
\psi_{n\bm{k}}^{\uparrow}\left(\bm{r}\right) & \psi_{n\bm{k}}^{\downarrow}\left(\bm{r}\right)\end{array}\right]^{\mathrm{T}}$, where $\mathrm{T}$ is the transpose operator. The electrons in
a solid environment have two different behaviors: those that are far
from the nuclei and {}``free''-like can be described by plane waves,
and those that are close to nuclei and unaffected by other nuclei
can be described by atomic like functions. Within FP-LAPW formalism,
the space is divided into two regions: a sphere with radius $R_{\alpha}$
around each atom, often called the muffin-tin region and the remaining
space is interstitial region.\cite{Singh1994,Blaha2001,Blugel2006}
As a result, the BFs of electrons are always divided into two parts.
Plane waves are used to construct the BFs in interstitial region

\begin{align}
\psi_{n\bm{k}}^{\sigma}(\bm{r}) & =\frac{1}{\sqrt{\Omega}}{\displaystyle \sum_{j}}z_{n\bm{k},j}^{\sigma}e^{i\left(\bm{k}+\bm{K}_{j}\right)\cdot\bm{r}},\;\bm{r}\in I,\label{eq:BFs_I}
\end{align}
where $\Omega$ is unit cell volume, $z_{n\bm{k},j}^{\sigma}$ is
the expansion coefficient, $\sigma$ and $n$ stand for spin and band
index, $\bm{k}$ for $\bm{k}$-points wave vector, $\bm{K}_{j}$ for
the $j$-th reciprocal-lattice vector, $j$ for the loop index of
every expansion term and up to a largest value by the condition $\left|\bm{k}+\bm{K}_{j}\right|\leq\bm{K}_{max}$,
and $\bm{K}_{max}$ for the cutoff vector. Within the muffin-tin region
(suppose the $\alpha$-th atom sphere with radius $R_{\alpha}$),
the BFs can be written as

\begin{multline}
\psi_{n\bm{k}}^{\sigma,\alpha}(\bm{r})={\displaystyle \sum_{lm}}\left[A_{lm}^{\sigma,\alpha}\left(n,\bm{k}\right)u_{l,1}^{\sigma,\alpha}+B_{lm}^{\sigma,\alpha}\left(n,\bm{k}\right)\dot{u}_{l,1}^{\sigma,\alpha}+C_{lm}^{\sigma,\alpha}\left(n,\bm{k}\right)u_{l,2}^{\sigma,\alpha}+D_{lm}^{\sigma,\alpha}\left(n,\bm{k}\right)u_{l,1/2}^{\sigma,\alpha}\right]Y_{lm}\left(\hat{\bm{r}}^{\alpha}\right),\\
\left|\bm{r}-\bm{\tau}^{\alpha}\right|\in\bm{R}_{\alpha},\label{eq:BFs_MT}
\end{multline}
 with

\begin{align}
A_{lm}^{\sigma,\alpha}\left(n,\bm{k}\right) & ={\displaystyle \sum_{j}}z_{n\bm{k},j}^{\sigma}\tilde{A}_{lm}^{\sigma,\alpha}\left(\bm{k}+\bm{K}_{j}\right)+{\displaystyle \sum_{j_{0}}}z_{n\bm{k},j_{0}}^{\sigma}\tilde{A}_{l_{0}m_{0}}^{\sigma,\alpha}\left(\bm{k}+\bm{K}_{j_{0}}\right)\delta_{l,l_{0}}\delta_{m,m_{0}},\nonumber \\
B_{lm}^{\sigma,\alpha}\left(n,\bm{k}\right) & ={\displaystyle \sum_{j}}z_{n\bm{k},j}^{\sigma}\tilde{B}_{lm}^{\sigma,\alpha}\left(\bm{k}+\bm{K}_{j}\right)+{\displaystyle \sum_{j_{0}}}z_{n\bm{k},j_{0}}^{\sigma}\tilde{B}_{l_{0}m_{0}}^{\sigma,\alpha}\left(\bm{k}+\bm{K}_{j_{0}}\right)\delta_{l,l_{0}}\delta_{m,m_{0}},\nonumber \\
C_{lm}^{\sigma,\alpha}\left(n,\bm{k}\right) & ={\displaystyle \sum_{j_{0}}}z_{n\bm{k},j_{0}}^{\sigma}\tilde{C}_{l_{0}m_{0}}^{\sigma,\alpha}\left(\bm{k}+\bm{K}_{j_{0}}\right)\delta_{l,l_{0}}\delta_{m,m_{0}},\nonumber \\
D_{lm}^{\sigma,\alpha}\left(n,\bm{k}\right) & ={\displaystyle \sum_{j_{0}}}z_{n\bm{k},j_{0}}^{\sigma}\tilde{D}_{l_{0}m_{0}}^{\sigma,\alpha}\left(\bm{k}+\bm{K}_{j_{0}}\right)\delta_{l,l_{0}}\delta_{m,m_{0}}.\label{eq:BFs_MT_coef}
\end{align}
 where $\bm{r}^{\alpha}=\bm{r}-\bm{\tau}^{\alpha}$ and $\bm{\tau}^{\alpha}$
is the position of atom $\alpha$; $lm$ is the angular momentum index;
$Y_{lm}$ is spherical harmonics. In above formulas, $u_{l,1}^{\sigma,\alpha}\equiv u_{l}^{\sigma}\left(r^{\alpha},E_{l,1}^{\alpha}\right)$
and $\dot{u}_{l,1}^{\sigma,\alpha}\equiv\dot{u}_{l}^{\sigma}\left(r^{\alpha},E_{l,1}^{\alpha}\right)$
are the radial solutions of scalar-relativistic Schr�dinger equation
of atom $\alpha$ and their energy derivatives, both evaluated at
energy $E_{l,1}^{\alpha}$. The local orbit radial functions $u_{l,2}^{\sigma,\alpha}\equiv u_{l}^{\sigma}\left(r^{\alpha},E_{l,2}^{\alpha}\right)$
are added to the $u_{l,1}^{\sigma,\alpha}$ and $\dot{u}_{l,1}^{\sigma,\alpha}$
for semi-core states (when $l=l_{0}$) and aimed to increase the variational
freedom of standard basis functions. The last radial functions $u_{l,1/2}^{\sigma,\alpha}\equiv u_{l}^{\sigma}\left(r^{\alpha},E_{l,1/2}^{\alpha}\right)$,
as the radial solution of full-relativistic Dirac equation, is also
added to the $u_{l}^{\sigma}$ and $\dot{u}_{l,1}^{\sigma,\alpha}$
but only for $5p_{1/2}$ or $6p_{1/2}$ orbits in heavy elements.\cite{Junes2001}
This extended full-relativistic local orbit can improve the accuracy
of second-variational step when taking account of SOC. The $\tilde{A}_{lm}^{\sigma,\alpha}$
and $\tilde{B}_{lm}^{\sigma,\alpha}$ are the coefficients of LAPW
basis set, and $\tilde{B}_{lm}^{\sigma,\alpha}$is zero when APW basis
set is used. $\tilde{A}_{l_{0}m_{0}}^{\sigma,\alpha}$, $\tilde{B}_{l_{0}m_{0}}^{\sigma,\alpha}$
, $\tilde{C}_{l_{0}m_{0}}^{\sigma,\alpha}$, and $\tilde{D}_{l_{0}m_{0}}^{\sigma,\alpha}$
are the coefficients of local orbit basis set. These coefficients
can be determined by imposing various boundary conditions at the muffin-tin
boundaries.\cite{Singh1994,Blaha2001,Blugel2006}

Considering a lattice division within BZ, the overlap matrix between
$\bm{k}$ point and its nearest-neighbor $\bm{k+b}$ has the form

\begin{eqnarray}
M_{mn}^{(\bm{k,b})} & = & \left\langle u_{m,\bm{k}}^{\uparrow}|u_{n,\bm{k+b}}^{\uparrow}\right\rangle +\left\langle u_{m,\bm{k}}^{\downarrow}|u_{n,\bm{k+b}}^{\downarrow}\right\rangle .\label{eq:Overlap_Matrix}
\end{eqnarray}
 Overlap matrix $M_{mn}^{(\bm{k,b})}$ is a very useful quantity in
many Berry-phase related calculations,\cite{King-Smith1993,Resta1994}
and the detailed formulas for its calculations are demonstrated in
Appendix.

\subsection{Parity criterion in centrosymmetric system\label{sub:II B}}

For systems with spatial inversion symmetry, $Z_{2}$ invariants can
be obtained by parity analysis developed by Fu and Kane\cite{Fu2007a}.
In 3D system there are eight time-reversal invariant momenta (TRIM)
in BZ, $\mathbf{\Gamma}_{i=\left(n_{1}n_{2}n_{3}\right)}=\frac{1}{2}\left(n_{1}\bm{G}_{1}+n_{2}\bm{G}_{2}+n_{3}\bm{G}_{3}\right)$,
where $\bm{G}_{j}$ are primitive reciprocal-lattice vectors with
$n_{j}=0\textrm{, or }1$. The $Z_{2}$ invariants are determined
by the quantities

\begin{equation}
\delta_{i}=\prod_{m=1}^{N}\xi_{2m}\left(\Gamma_{i}\right).\label{eq:delta_i}
\end{equation}
 Here, $\xi_{2m}\left(\Gamma_{i}\right)=\pm1$ is the parity eigenvalue
of the \textit{2m}-th occupied energy band at TRIMs $\Gamma_{i}$,
i.e. $\left\langle \Psi_{2m,\Gamma_{i}}\left|P\right|\Psi_{2m,\Gamma_{i}}\right\rangle $,
where $P$ is parity operator. Because of the Kramers degeneracy at
TRIMs, the \textit{2m}-th and (\textit{2m-1})-th occupied bands have
the same eigenvalues, i.e., $\xi_{2m}=\xi_{2m-1}$. In 3D system,
there are four independent invariants $\nu_{0};(\nu_{1}\nu_{2}\nu_{3})$,
given by\cite{Fu2007a}

\begin{equation}
\left(-1\right)^{\nu_{0}}=\prod_{i=1}^{8}\delta_{i},\label{eq:Z2_v0}
\end{equation}

\begin{equation}
\left(-1\right)^{\nu_{k}}=\prod_{n_{k}=1,n_{j\neq k}=0,1}\delta_{i=\left(n_{1}n_{2}n_{3}\right)},\label{eq:Z2_vk}
\end{equation}
 where $\nu_{0}$ is independent of the choice of primitive reciprocal-lattice
vectors $\bm{G}_{j}$ while $\nu_{1}$, $\nu_{2}$, and $\nu_{3}$
are not. A nonzero $\nu_{0}$ indicates that the system is a strong
topological insulator (STI). When $\nu_{0}=0$, the systems are further
classified according to $\nu_{1}$, $\nu_{2}$ , and $\nu_{3}$. The
systems with $\nu_{1,2,\textrm{or }3}\neq0$ are called weak topological
insulators (WTI), while $0;(000)$ is normal insulator (NI).

To obtain $Z_{2}$ invariants, the basic job is to calculate the matrix
elements of parity operator $\left\langle \Psi_{n\bm{k}}\left(\bm{r}\right)\left|P\right|\Psi_{n\bm{k}}\left(\bm{r}\right)\right\rangle $
with even band index $n$ at eight TRIMs $\Gamma_{i}$. The parity
operator $P$ is defined as $\left\{ I;\bm{t}\right\} $, where $I$
is an inverse matrix making $\bm{r}\rightarrow-\bm{r}$ and $\bm{t}$
is a translational vector. Since parity operation will not change
spin component of BFs, then,

\begin{equation}
\left\langle \Psi_{n\bm{k}}\left(\bm{r}\right)\left|P\right|\Psi_{n\bm{k}}\left(\bm{r}\right)\right\rangle =\left\langle \psi_{n\bm{k}}^{\uparrow}\left(\bm{r}\right)\left|P\right|\psi_{n\bm{k}}^{\uparrow}\left(\bm{r}\right)\right\rangle +\left\langle \psi_{n\bm{k}}^{\downarrow}\left(\bm{r}\right)\left|P\right|\psi_{n\bm{k}}^{\downarrow}\left(\bm{r}\right)\right\rangle .
\end{equation}

In the following, we take $\left\langle \psi_{n\bm{k}}^{\uparrow}\left(\bm{r}\right)\left|P\right|\psi_{n\bm{k}}^{\uparrow}\left(\bm{r}\right)\right\rangle $
as an example and suppress the spin index from here. Suppose that
$\tilde{\psi}_{n\bm{k}}\left(\bm{r}\right)=P\psi_{n\bm{k}}\left(\bm{r}\right)$
and inversion center at $\frac{\bm{t}}{2}$, then $\tilde{\psi}_{n\bm{k}}\left(\frac{\bm{t}}{2}-\bm{r}\right)=\psi_{n\bm{k}}\left(\frac{\bm{t}}{2}+\bm{r}\right)$.
It can be rewritten as $\tilde{\psi}_{n\bm{k}}\left(\bm{r}\right)=\psi_{n\bm{k}}\left(\bm{t}-\bm{r}\right)$,
and finally we have $P\psi_{n\bm{k}}\left(\bm{r}\right)=\psi_{n\bm{k}}\left(\bm{t}-\bm{r}\right)$.
The matrix elements of parity operator are divided into two parts

\begin{equation}
\left\langle \psi_{n\bm{k}}\left(\bm{r}\right)\left|P\right|\psi_{n\bm{k}}\left(\bm{r}\right)\right\rangle =\left\langle \psi_{n\bm{k}}\left(\bm{r}\right)\left|P\right|\psi_{n\bm{k}}\left(\bm{r}\right)\right\rangle _{I}+{\displaystyle \sum_{\alpha}}\left\langle \psi_{n\bm{k}}^{\alpha}\left(\bm{r}\right)\left|P\right|\psi_{n\bm{k}}^{\alpha}\left(\bm{r}\right)\right\rangle _{MT^{\alpha}}.
\end{equation}
 The contribution of interstitial region is

\begin{align}
\left\langle \psi_{n\bm{k}}\left(\bm{r}\right)\left|P\right|\psi_{n\bm{k}}\left(\bm{r}\right)\right\rangle _{I} & =\frac{1}{\Omega}\sum_{ij}z_{n\bm{k},i}^{*}z_{n\bm{k},j}\int_{cell}e^{-i\left(\bm{k}+\bm{K}_{i}\right)\cdot\mathbf{r}}e^{i\left(\bm{k}+\bm{K}_{j}\right)\cdot\left(\bm{t}-\bm{r}\right)}\Delta\left(\bm{r}\right)d^{3}r\nonumber \\
 & =\frac{1}{\Omega}\sum_{ij}z_{n\bm{k},i}^{*}z_{n\bm{k},j}e^{i\left(\bm{k}+\bm{K}_{j}\right)\cdot\bm{t}}\Delta\left(2\bm{k}+\bm{K}_{i}+\bm{K}_{j}\right).
\end{align}
 Here, $\Delta(\bm{r})$ is a step function with zero value in muffin-tin
sphere and unit value in interstitial region and $\Delta(\bm{K})$
is its Fourier transformation. While inside the muffin-tin region,
the radial coefficients in Eq. (\ref{eq:BFs_MT_coef}) can be rewritten
as a product of two parts, one of which depends on atomic positions
and the other does not. For example, $A_{lm}^{\alpha}\left(n,\bm{k}\right)=\sum_{j}\eta_{n,lm}\left(\bm{k}+\bm{K}_{j},R_{\alpha}\right)e^{i\left(\bm{k}+\bm{K}_{j}\right)\cdot\bm{\tau}^{\alpha}}$,
where $\boldsymbol{\tau}^{\alpha}$ is the position of $\alpha$-th
atom and $\eta_{n,lm}\left(\mathbf{k}+\bm{K}_{j},R_{\alpha}\right)$
is independent of $\boldsymbol{\tau}^{\alpha}$. Therefore,

\begin{equation}
PA_{lm}^{\alpha}\left(n,\bm{k}\right)={\displaystyle \sum_{j}}\eta_{n,lm}\left(\bm{k}+\bm{K}_{j},R_{\alpha}\right)e^{i\left(\bm{k}+\bm{K}_{j}\right)\cdot\left(\bm{t}-\boldsymbol{\tau}^{\alpha}\right)}.
\end{equation}
 If atom $\alpha$ is operated by parity operator, it must overlap
another equivalent atom $\beta$ by translated integer numbers of
primitive real-lattice, i.e. $\bm{t}-\bm{\tau}^{\alpha}=\bm{R}_{h}+\bm{\tau}^{\beta}$
with $\bm{R}_{h}=h_{1}\mathbf{a}_{1}+h_{2}\mathbf{a}_{2}+h_{3}\mathbf{a}_{3}$,
where $h_{j}$ is integer number and $\mathbf{a}_{j}$ is primitive
real-lattice vector. Then above equation can be rewritten as

\begin{align}
PA_{lm}^{\alpha}\left(n,\bm{k}\right) & ={\displaystyle \sum_{j}}\eta_{n,lm}\left(\bm{k}+\bm{K}_{j},R_{\beta}\right)e^{i\left(\bm{k}+\bm{K}_{j}\right)\cdot\bm{\tau}^{\beta}}e^{i\bm{k}\cdot\bm{R}_{h}}\nonumber \\
 & =A_{lm}^{\beta}\left(n,\bm{k}\right)e^{i\bm{k}\cdot\bm{R}_{h}},
\end{align}
 and there are similar operations for $PB_{lm}^{\alpha}\left(n,\bm{k}\right)$,
$PC_{lm}^{\alpha}\left(n,\bm{k}\right)$, and $PD_{lm}^{\alpha}\left(n,\bm{k}\right)$.
The spherical harmonics operated by parity operator is, $PY_{lm}\left(\hat{\bm{r}}^{\alpha}\right)=\left(-1\right)^{l}Y_{lm}\left(\hat{\bm{r}}^{\alpha}\right)$.
Then we have

\begin{equation}
P\psi_{n\bm{k}}^{\alpha}(\bm{r})={\displaystyle e^{i\bm{k}\cdot\bm{R}_{h}}\sum_{lm}}\left[A_{lm}^{\beta}\left(n,\bm{k}\right)u_{l,1}^{\alpha}+B_{lm}^{\beta}\left(n,\bm{k}\right)\dot{u}_{l,1}^{\alpha}+C_{lm}^{\beta}\left(n,\bm{k}\right)u_{l,2}^{\alpha}+D_{lm}^{\beta}\left(n,\bm{k}\right)u_{l,1/2}^{\alpha}\right]\left(-1\right)^{l}Y_{lm}\left(\hat{\bm{r}}^{\alpha}\right).
\end{equation}
 Therefore, we can easily obtain $P\psi_{n\bm{k}}^{\alpha}(\bm{r})$
by using radial coefficients of atom $\beta$, $A_{lm}^{\beta}\left(n,\bm{k}\right)$,
$B_{lm}^{\beta}\left(n,\bm{k}\right)$, $C_{lm}^{\beta}\left(n,\bm{k}\right)$,
and $D_{lm}^{\beta}\left(n,\bm{k}\right)$, which have already been
calculated. Finally, the calculation of $\left\langle \psi_{n\bm{k}}^{\alpha}\left(\bm{r}\right)\left|P\right|\psi_{n\bm{k}}^{\alpha}\left(\bm{r}\right)\right\rangle $
is very similar to $\left\langle \psi_{n\bm{k}}^{\alpha}\left(\bm{r}\right)|\psi_{n\bm{k}}^{\alpha}\left(\bm{r}\right)\right\rangle $,
which can be found in Appendix.

\subsection{Lattice calculation of $Z_{2}$ invariants in noncentrosymmetric
system\label{sub:II C}}

A nontrivial topological invariant $Z_{2}$ can be interpreted as
an obstruction to make the BFs smoothly defined over BZ under time-reversal
constrains. \cite{Fu2007b,Moore2007,Roy2009} Here, we present a lattice
evaluation of the $Z_{2}$ invariants in terms of the Berry gauge
potential and Berry curvature associated with the BFs.\cite{Fukui2007}
This method has been recently applied to our first-principles studies
of ternary half-Heusler\cite{Xiao2010a} and chalcopyrite\cite{Feng2011}
TIs and QSHE in Silicene thinfilm.\cite{Liu2011}

\begin{figure}
\includegraphics[width=3.5in]{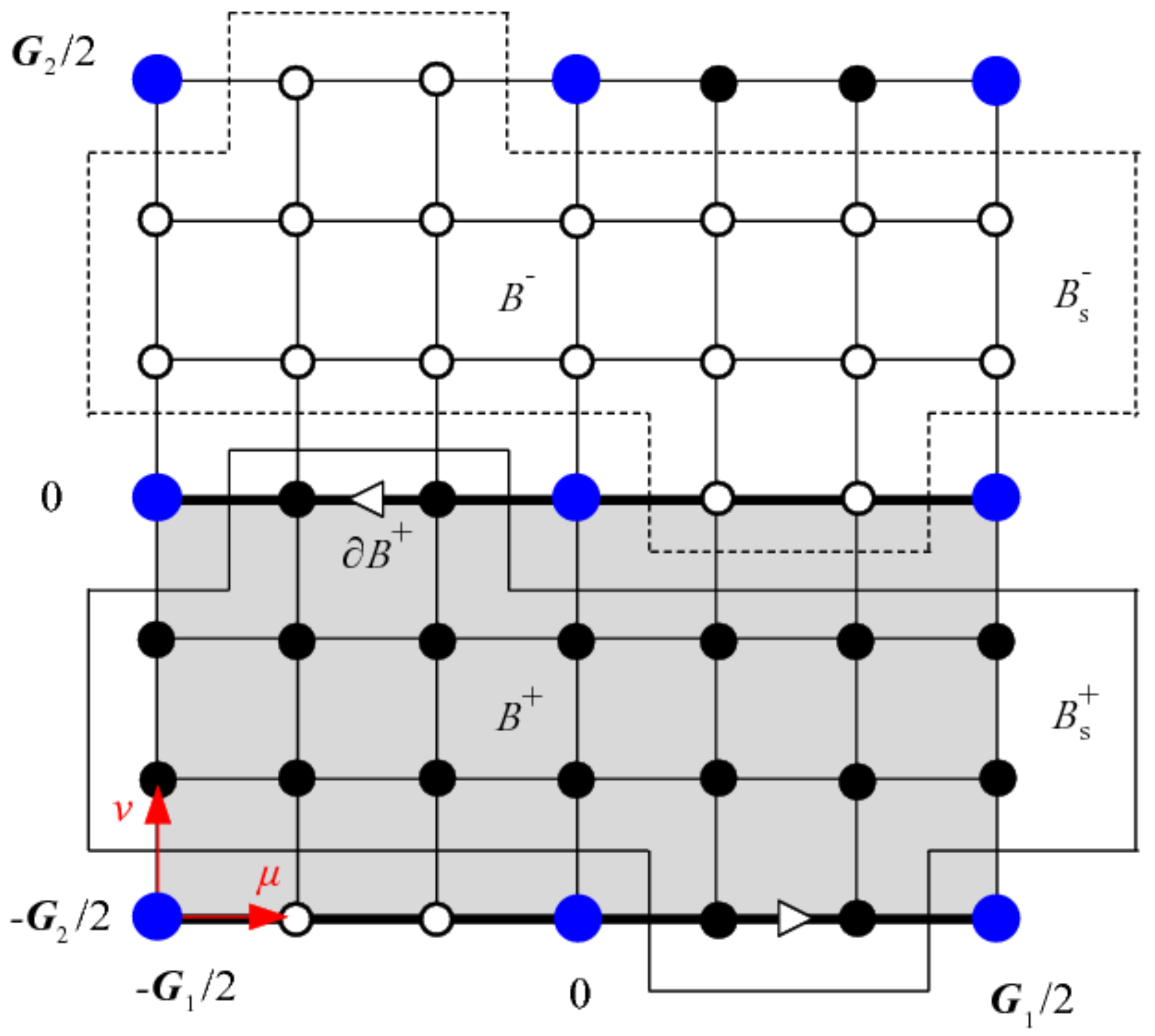}\caption{(Color online) Schematic drawing of lattice mesh in a two-dimensional
Brillouin zone. Under the time-reversal constraint, only \emph{half}
of Brillouin zone $\mathcal{B}^{+}$ is needed, which is denoted by
shaded region. The thick lines indicate the boundary of $\mathcal{B}^{+}$,
i.e., $\partial\mathcal{B}^{+}$, and the open arrows denote their
directions. All $\bm{k}$-points are divided into three classes: $\mathcal{B}_{s}^{+}$,
$\mathcal{B}_{s}^{-}$, and $\mathcal{B}_{s}^{0}$, which are represented
by small (black) solid, small (black) open and large (blue) solid
circles, respectively.}

\label{Flo:BZ_2D} 
\end{figure}

We first briefly describe the formalism for a 2D system. It was shown
by Fu and Kane\cite{Fu2006} that under the time-reversal constraint,
the $Z_{2}$ invariants can be written as

\begin{equation}
Z_{2}=\frac{1}{2\pi}\left[\oint_{\partial\mathcal{B}^{+}}d\bm{k}\cdot\bm{\mathcal{A}}\left(\bm{k}\right)-\int_{\mathcal{B}^{+}}d^{2}k\,\mathcal{F}\left(\bm{k}\right)\right]\textrm{ mod 2},\label{eq:Z2}
\end{equation}
 where $\mathcal{B}^{+}$ and $\partial\mathcal{B}^{+}$ represent
\emph{half} of BZ and its boundary (Fig. \ref{Flo:BZ_2D}). The central
quantities are the Berry connection 
\begin{equation}
\bm{\mathcal{A}}=i\sum_{n}\left\langle u_{n}\left(\bm{k}\right)\mid\bm{\nabla}_{\bm{k}}u_{n}\left(\bm{k}\right)\right\rangle \label{eq:Berry_conn}
\end{equation}
 and the Berry curvature

\begin{equation}
\mathcal{F}\left(\bm{k}\right)=\bm{\nabla}_{\bm{k}}\times\bm{\mathcal{A}}\left(\bm{k}\right)\mid_{z},\label{eq:Berry_curv}
\end{equation}
 where $\left|u_{n}\left(\bm{k}\right)\right\rangle $ is the periodic
part of BFs and the sum is over occupied bands. TIs are characterized
by $Z_{2}=1$ while normal insulators have $Z_{2}=0$.

In the following, we introduce the calculation of $\left|u_{n}\left(\bm{k}\right)\right\rangle $
in \emph{half} of BZ referred to as $\mathcal{B}^{+}$ ($\left[-\bm{G}_{1}/2,\bm{G}_{1}/2\right]\otimes\left[-\bm{G}_{2}/2,0\right]$)
according to the time-reversal constraint. As shown in Fig. \ref{Flo:BZ_2D},
the $\bm{k}$-points on a 2D BZ with $N\times N$ division are divided
into three classes: $\mathcal{B}_{s}^{+}$, $\mathcal{B}_{s}^{-}$,
and $\mathcal{B}_{s}^{0}$. Firstly, we obtain $\left|u_{n}\left(\bm{k}\right)\right\rangle $
in $\mathcal{B}_{s}^{+}$ except for the points on the right edge.
The points on the right edge ($\bm{k}'=\bm{k}+\bm{G_{1}}$) are the
periodic images of those on the left edge ($\bm{k}$), and can be
calculated by using the periodic gauge\cite{King-Smith1993,Resta1994}

\begin{equation}
\left|u_{n}\left(\bm{k}+\bm{G_{1}}\right)\right\rangle =e^{-i\bm{G_{1}\cdot r}}\left|u_{n}\left(\bm{k}\right)\right\rangle .\label{eq:Peri_gauge}
\end{equation}
 Secondly, we consider the $\mathcal{B}_{s}^{-}$ points on the boundary
$\partial\mathcal{B}^{+}$, i.e., the left part of the bottom edge
and the right part of top edge. These points $-\bm{k}\in\mathcal{B}_{s}^{-}$
are the Kramers doublets of $\bm{k}\in\mathcal{B}_{s}^{+}$ points,
so they can be calculated by the time-reversal constraint,

\begin{equation}
\begin{array}{cc}
\left|u_{n}\left(-\bm{k}\right)\right\rangle =\Theta\left|u_{n}\left(\bm{k}\right)\right\rangle , & \textrm{for}\end{array}\bm{k}\in\mathcal{B}_{s}^{+}.\label{eq:Time_con1}
\end{equation}
where $\Theta=-i\sigma_{y}K$ is the time-reversal operator with $K$
the complex conjugation. Note that translational phase factors must
be properly considered. For example, $\bm{k}'\in\mathcal{B}_{s}^{-}$
and $\bm{k}\in\mathcal{B}_{s}^{+}$ are two points which are centrosymmetric
about the midpoint of the bottom edge, i.e., $\bm{k}'=-\bm{k}-\bm{G}_{2}$,
then we have 
\begin{eqnarray}
\left|u_{n}\left(\bm{k}'\right)\right\rangle  & = & \left|u_{n}\left(-\bm{k-\bm{G}_{2}}\right)\right\rangle \nonumber \\
 & = & e^{i\bm{G_{2}\cdot r}}\left|u_{n}\left(-\bm{k}\right)\right\rangle \nonumber \\
 & = & e^{i\bm{G_{2}\cdot r}}\Theta\left|u_{n}\left(\bm{k}\right)\right\rangle .
\end{eqnarray}
 Finally, we calculate $\left|u_{n}\left(\bm{k}\right)\right\rangle $
on TRIMs, i.e., $\mathcal{B}_{s}^{0}$, satisfied by $\Theta H(\bm{k})\Theta^{-1}=H(\bm{k})$.
The eigenvalues are $\ldots\varepsilon_{2n-1}\left(\bm{k}\right)=\varepsilon_{2n}\left(\bm{k}\right)\leq\varepsilon_{2n+1}\left(\bm{k}\right)=\varepsilon_{2n+2}\left(\bm{k}\right)\ldots$
because of the Kramers degeneracy. In this situation, the time-reversal
constraint is given by 
\begin{equation}
\begin{array}{cc}
\left|u_{2n}\left(-\bm{k}\right)\right\rangle =\Theta\left|u_{2n-1}\left(\bm{k}\right)\right\rangle , & -\bm{k}\textrm{ and }\bm{k}\in\mathcal{B}_{s}^{0}\end{array}.\label{eq:Time_con2}
\end{equation}
 There are six TRIMs in \emph{half} of BZ $\mathcal{B}^{+}$, $-\bm{G}_{1}/2-\bm{G}_{2}/2$,
$-\bm{G}_{1}/2$, $-\bm{G}_{2}/2$, $\bm{0}$, $\bm{G}_{1}/2-\bm{G}_{2}/2$,
and $\bm{G}_{1}/2$. For the former four points, the $2n$-th eigenstates
can be obtained from $(2n-1)$-th eigenstates by using above constraint.
Here, one should also consider the translational phase factor, for
example,

\begin{eqnarray}
\left|u_{2n}\left(-\bm{G}_{1}/2-\bm{G}_{2}/2\right)\right\rangle  & = & e^{i(\bm{G}_{1}+\bm{G}_{2})\bm{\cdot r}}\left|u_{2n}\left(\bm{G}_{1}/2+\bm{G}_{2}/2\right)\right\rangle \nonumber \\
 & = & e^{i(\bm{G}_{1}+\bm{G}_{2})\bm{\cdot r}}\Theta\left|u_{2n-1}\left(-\bm{G}_{1}/2-\bm{G}_{2}/2\right)\right\rangle .
\end{eqnarray}
 The other two points, $\bm{G}_{1}/2-\bm{G}_{2}/2$, and $\bm{G}_{1}/2$,
can be obtained by their periodic image points, i.e. $\left|u_{n}\left(\bm{G}_{1}/2-\bm{G}_{2}/2\right)\right\rangle =e^{-i\bm{G}_{1}\cdot\bm{r}}\left|u_{n}\left(-\bm{G}_{1}/2-\bm{G}_{2}/2\right)\right\rangle $,
$\left|u_{n}\left(\bm{G}_{1}/2\right)\right\rangle =e^{-i\bm{G}_{1}\cdot\bm{r}}\left|u_{n}\left(-\bm{G}_{1}/2\right)\right\rangle $.

After applying the time-reversal constrain Eq. (\ref{eq:Time_con1})
and Eq. (\ref{eq:Time_con2}) and periodic gauge Eq. (\ref{eq:Peri_gauge}),
we have obtained a new set of basis functions $\left|\tilde{u}_{n}\left(\bm{k}\right)\right\rangle $.
Next, we introduce the link variable that is central to many Berry-phase
related calculations,\cite{King-Smith1993,Resta1994} given by

\begin{equation}
U_{\bm{\mu}}\left(\bm{k}_{j}\right)=N_{\bm{\mu}}^{-1}\left(\bm{k}_{j}\right)\det\left\langle \tilde{u}_{m}\left(\bm{k}_{j}\right)\mid\tilde{u}_{n}\left(\bm{k}_{j}+\bm{\mu}\right)\right\rangle ,
\end{equation}
 where $N_{\bm{\mu}}^{-1}\left(\bm{k}_{j}\right)=\left|\det\left\langle \tilde{u}_{m}\left(\bm{k}_{j}\right)\mid\tilde{u}_{n}\left(\bm{k}_{j}+\bm{\mu}\right)\right\rangle \right|$
is the normalizing factor and $\bm{\mu}$ is the unit vector on the
$\bm{k}$-mesh. In practice, $\left\langle \tilde{u}_{m}\left(\bm{k}_{j}\right)\mid\tilde{u}_{n}\left(\bm{k}_{j}+\bm{\mu}\right)\right\rangle $
is the overlap matrix $\left\langle u_{m,\bm{k}}|u_{n,\bm{k+\mu}}\right\rangle $
or its derivatives with the time-reversal operator $\Theta$ including
$\left\langle u_{m,\bm{k}}|\Theta u_{n,\bm{k+\mu}}\right\rangle $,
$\left\langle \Theta u_{m,\bm{k}}|u_{n,\bm{k+\mu}}\right\rangle $,
and $\left\langle \Theta u_{m,\bm{k}}|\Theta u_{n,\bm{k+\mu}}\right\rangle $.
The calculation of $\left\langle \tilde{u}_{m}\left(\bm{k}_{j}\right)\mid\tilde{u}_{n}\left(\bm{k}_{j}+\bm{\mu}\right)\right\rangle $
is demonstrated in Appendix.

The finite element expressions for Berry connection $\bm{\mathcal{A}}$
and Berry curvature $\mathcal{F}$ are

\begin{equation}
\bm{\mathcal{A}}_{\bm{\mu}}\left(\bm{k}_{j}\right)=\textrm{Im}\log U_{\bm{\mu}}\left(\bm{k}_{j}\right),
\end{equation}
 and

\begin{equation}
\mathcal{F}\left(\bm{k}_{j}\right)=\textrm{Im}\log U_{\bm{\mu}}\left(\bm{k}_{j}\right)U_{\bm{\nu}}\left(\bm{k}_{j}+\bm{\mu}\right)U_{\bm{\mu}}^{-1}\left(\bm{k}_{j}+\bm{\nu}\right)U_{\bm{\nu}}^{-1}\left(\bm{k}_{j}\right),
\end{equation}
 where the return value of the complex logarithm function is confined
to its principal branch $(-\pi,\pi]$. We can then insert these expressions
into Eq. (\ref{eq:Z2}) to calculate the $Z_{2}$ invariants.

To visualize the above procedure, an integer field $n(\bm{k}_{j})$
can be defined for each torus:

\begin{equation}
n(\bm{k}_{j})=\frac{1}{2\pi}\left\{ \left[\Delta_{\bm{\nu}}\bm{\mathcal{A}}_{\bm{\mu}}\left(\bm{k}_{j}\right)-\Delta_{\bm{\mu}}\bm{\mathcal{A}}_{\bm{\nu}}\left(\bm{k}_{j}\right)\right]-\mathcal{F}\left(\bm{k}_{j}\right)\right\} ,
\end{equation}
 where $\Delta_{\bm{\mu}}$ is the forward difference operator. The
$Z_{2}$ invariants are given by the sum of the $n$-field in \emph{half}
of the BZ, i.e., $Z_{2}=\sum_{\bm{k}_{j}\in\mathcal{B}^{+}}n(\bm{k}_{j})\textrm{ mod 2}$.
The sum of $n$-field configuration over the entire BZ gives a vanished
Chern number for time-reversal invariant systems. It must be emphasized
that the $n$-field summed over \emph{half} of BZ is gauge-invariant
module 2 even though itself depends on a specific gauge choice.

\begin{figure}
\includegraphics[width=3.5in]{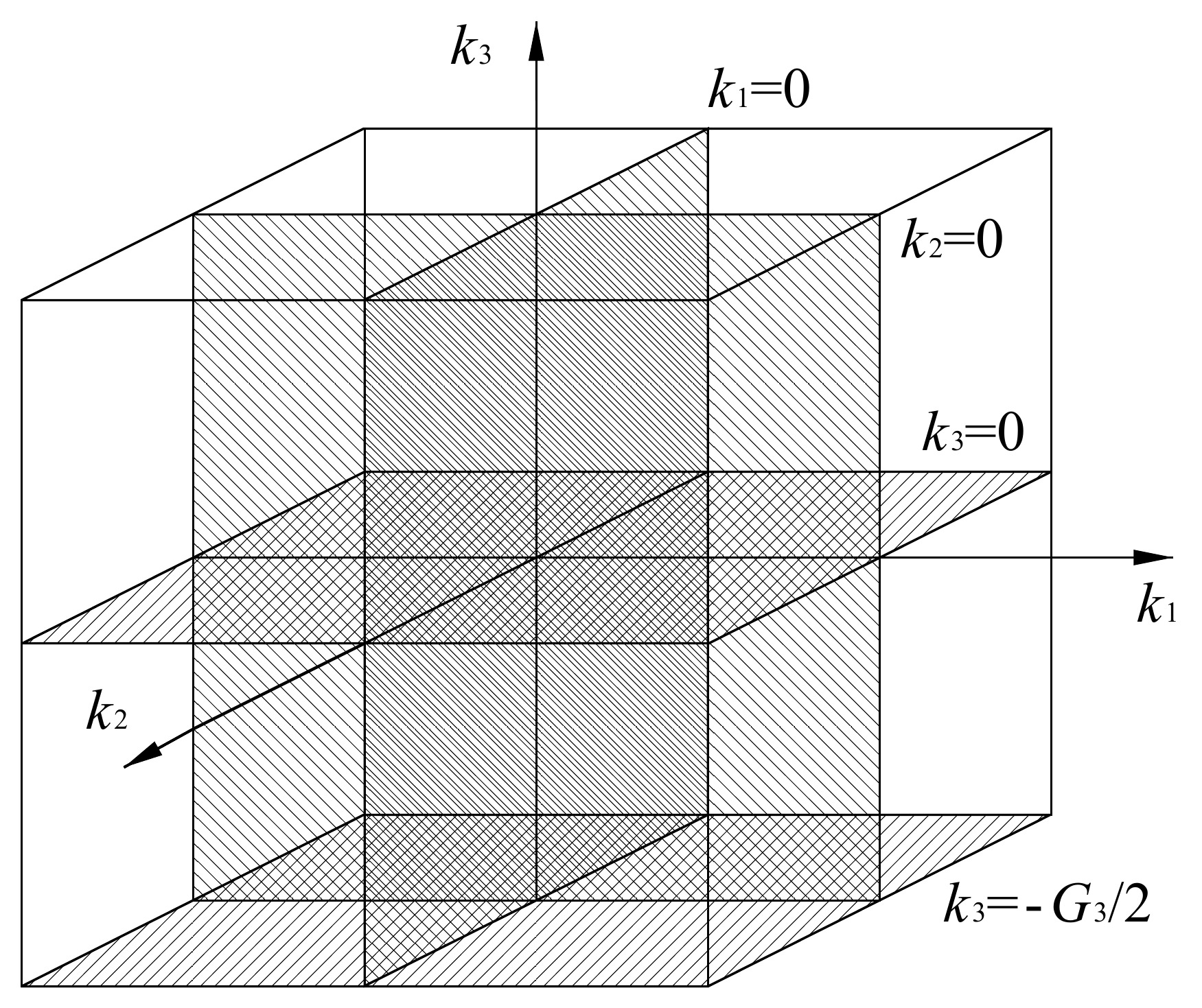}

\caption{Schematic drawing of four independent tori in a three-dimensional
Brillouin zone. The four independent tori $T(Z_{0})$, $T(Z_{1})$,
$T(X_{0})$ and $T(Y_{0})$ are located at $k_{3}=0$, $k_{3}=-G_{3}/2$,
$k_{1}=0$, and $k_{2}=0$, respectively.}

\label{Flo:BZ_3D} 
\end{figure}

In 3D system, there are six possible 2D tori. These 2D tori are defined
as follows: for example, the torus $T(X_{0})$ is spanned by $G_{2}$
and $G_{3}$ with the first component fixed at 0, and $T(X_{1})$
is obtained by fixing the first component at $-G_{1}/2$. The other
four tori $T(Y_{0})$, $T(Y_{1})$, $T(Z_{0})$, and $T(Z_{1})$ are
defined similarly. For each torus, one can calculate the corresponding
$Z_{2}$ invariants, $x_{0}$, $x_{1}$, $y_{0}$, $y_{1}$, $z_{0}$,
and $z_{1}$, by using the steps outlined above for 2D BZ. Out of
the six possible $Z_{2}$ invariant only four of them are independent
due to the constraint $x_{0}+x_{1}=y_{0}+y_{1}=z_{0}+z_{1}\textrm{ (mod 2)}$.
Following Refs. \onlinecite{Fu2007b,Moore2007,Roy2009}, we denote
four independent $Z_{2}$ invariants by $\nu_{0};(\nu_{1}\nu_{2}\nu_{3})$,
with $\nu_{0}=(z_{0}+z_{1})\textrm{ mod 2}$, $\nu_{1}=x_{1}$, $\nu_{2}=y_{1}$
and $\nu_{3}=z_{1}$. The corresponding four independent tori $T(Z_{0})$,
$T(Z_{1})$, $T(X_{0})$ and $T(Y_{0})$ are shown in Fig. \ref{Flo:BZ_3D}.

\section{results}

In this section, we apply our methods to both centrosymmetric and
noncentrosymmetric systems. In the case of centrosymmetric compounds
Bi$_{2}$Se$_{3}$ and Sb$_{2}$Se$_{3}$, our parity analysis shows
that Bi$_{2}$Se$_{3}$ is a STI while Sb$_{2}$Se$_{3}$ is a NI.
The lattice calculation of $Z_{2}$ invariants has also been used
as a double check and the results are consistent with the parity analysis.
We then turn to noncentrosymmetric compounds LuPtBi, AuTlS$_{2}$
and CdSnAs$_{2}$. By turning lattice constant, we studied three different
topological phases of LuPtBi, i.e., STI, topological metal (TM), and
NI. Furthermore, the $Z_{2}$ invariants show that chalcopyrite compounds
AuTlS$_{2}$ and CdSnAs$_{2}$ are STI and NI, respectively, in their
native states without any strain.

The calculations of band structures and $Z_{2}$ invariants in this
work were performed using FP-LAPW method,\cite{Singh1994,Blugel2006}
implemented in the package \textsc{wien2k}.\cite{Blaha2001} We used
two types of exchange-correlation potentials. The generalized gradient
approximation (GGA)\cite{Perdew1996} was used for Bi$_{2}$Se$_{3}$
and Sb$_{2}$Se$_{3}$, while the modified Becke-Johnson exchange
potential together with local-density approximation for the correlation
potential (MBJLDA)\cite{Tran2009} was used for LuPtBi, AuTlS$_{2}$,
and CdSnAs$_{2}$ because the resulting band topology is sensitive
to the choice of exchange-correlation potentials in these systems.\cite{Feng2010}
The converged ground state was obtained using $K_{max}R_{MT}=9.0$
for each system, where $K_{max}$ is the maximum size of reciprocal-lattice
vector and $R_{MT}$ represents the smallest muffin-tin radius. The
$\bm{k}$-points sampling in BZ was also carefully checked such that
self-consistent field calculations were well converged. Spin-orbit
coupling was included by a second-variational procedure,\cite{Singh1994}
where states up to 9 Ry above Fermi level were included in the basis
expansion, and the relativistic $p_{1/2}$ corrections\cite{Junes2001}
were also considered for $5p$ and $6p$ orbit in order to improve
the accuracy for systems including heavy elements.

For a given system, the time taken by calculating of $Z_{2}$ invariants
depends on numbers of lattice divisions on four independent tori in
3D BZ and numbers of occupied bands considered below the Fermi level.
For most of systems, a $10\times10$ lattice division on each torus
is enough for obtaining a converged result just as mentioned in Ref.
\onlinecite{Fukui2007}. However, one must be very careful with
the cases of small local band gaps, for example the system shown in
Fig. \ref{Flo:Bands_LuPtBi}(c), $50\times50$ lattice division is
need to reach the convergence. The included number of occupied bands
should always been explicitly separated with other low-lying bands
with an obvious global energy gap. The principle is that these low-lying
bands are usually closed shell with much lower energy and should have
trivial band topology. In the following, we chose 18, 18, 30, 40 and
20 occupied bands for Bi$_{2}$Se$_{3}$, Sb$_{2}$Se$_{3}$, LuPtBi,
AuTlS$_{2}$ and CdSnAs$_{2}$, respectively.

\subsection{Centrosymmetric systems}

\begin{figure}
\includegraphics[width=3.5in]{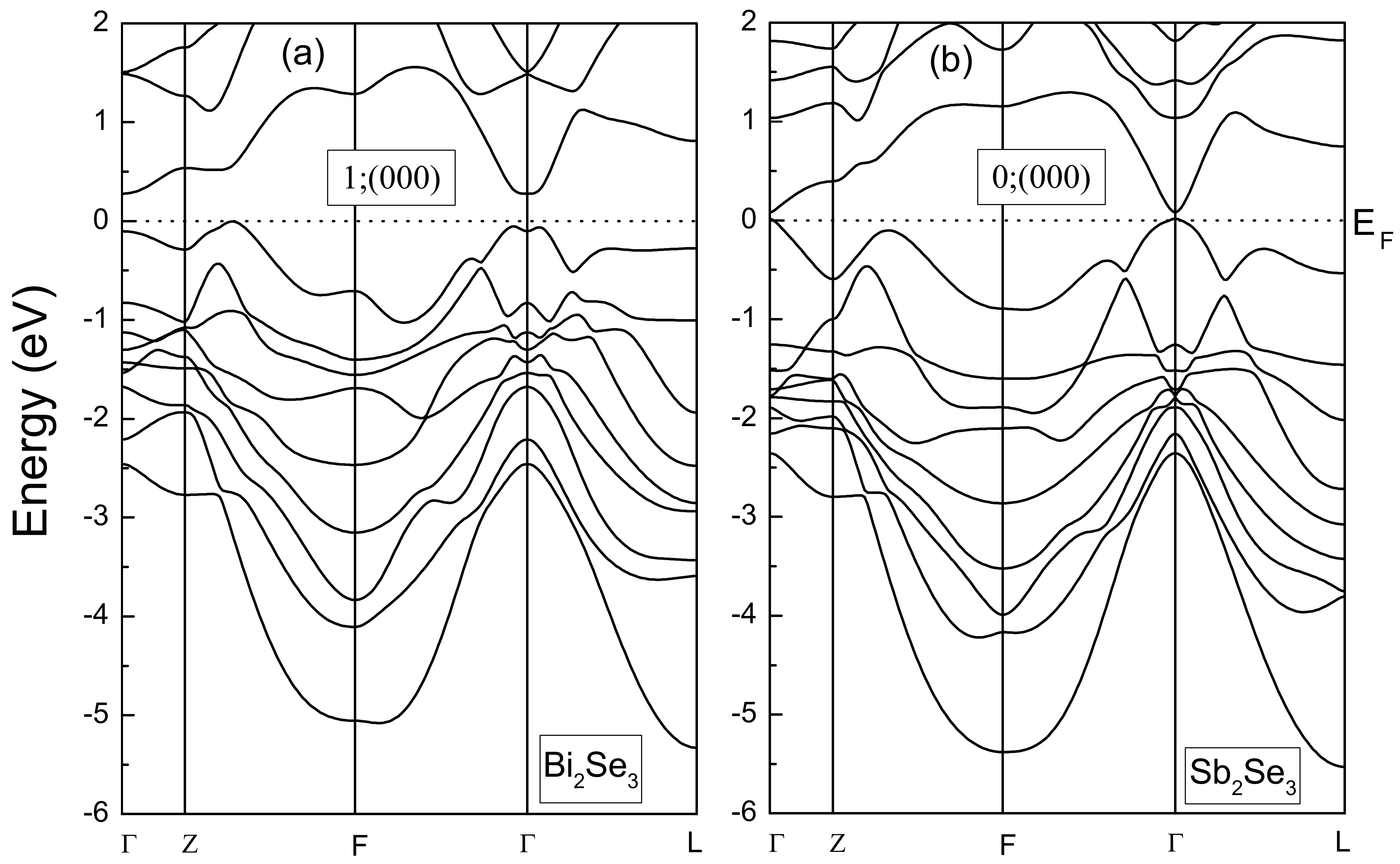}

\caption{Band structures of strong topological insulator Bi$_{2}$Se$_{3}$
with $Z_{2}$ invariants $1;(000)$ and normal insulator Sb$_{2}$Se$_{3}$
with $Z_{2}$ invariants $0;(000)$. The eighteen occupied bands (every
two of them are twofold degenerate) from $-6$ to $0$ eV are used
to calculate $Z_{2}$ invariants. The high-symmetry points in Brillouin
zone are the same as Ref. \onlinecite{Zhang2009}. }

\label{Flo:Bands_tetra} 
\end{figure}

To demonstrate the quality of our methods, we first test the centrosymmetric
systems Bi$_{2}$Se$_{3}$ and Sb$_{2}$Se$_{3}$. Recently, Bi$_{2}$Se$_{3}$
family of compounds have been both theoretically and experimentally
observed to be TIs with an exception of Sb$_{2}$Se$_{3}$.\cite{Zhang2009,Xia2009,Chen2009}
Tetradymite semiconductor Bi$_{2}$Se$_{3}$ family has a rhombohedral
crystal structure with space group $R\bar{3}m$ (No. 166) and three
nonequivalent atoms in a primitive cell. The calculated band structures
of Bi$_{2}$Se$_{3}$ and Sb$_{2}$Se$_{3}$ are presented in Fig.
\ref{Flo:Bands_tetra} with the lattice constants taken from previous
studies.\cite{Zhang2009} The 18 occupied bands ($-6\sim0$ eV) are
isolated from other low-lying bands and fully determine the topological
nature of the systems, so we consider them as a bands group in the
following calculation of $Z_{2}$ invariants.

Because the existence of spatial inversion symmetry, the parity criterion\cite{Fu2007a}
is applicable here. As a first step, we choose eight TRIMs in 3D BZ
with relative coordinates (0, 0, 0), (0, 0, 0.5), (0, 0.5, 0), (0,
0.5, 0.5), (0.5, 0, 0), (0.5, 0, 0.5), (0.5, 0.5, 0), (0.5, 0.5, 0.5)
in a primitive reciprocal-lattice. Then, we calculate the parity eigenvalues
of 9 occupied bands with even band index (sorted by energy) out of
18 occupied bands at every TRIM. The parity of each TRIM, $\delta_{i=1,2,...8}$
in Eq. (\ref{eq:delta_i}), are obtained by multiplying over the parity
eigenvalues of these 9 bands. The $Z_{2}$ invariant $\nu_{0}$ is
obtained by multiplying over the parities of all TRIMs according to
Eq. (\ref{eq:Z2_v0}), while $\nu_{k=1,2,3}$ by multiplying over
the parities of TRIMs resided in the same plane according to Eq. (\ref{eq:Z2_vk}).
The $\delta_{i}$ and $Z_{2}$ invariants are listed in Table \ref{Flo:Parity_TRIMs}.
The $Z_{2}$ invariants are $1;(000)$ for Bi$_{2}$Se$_{3}$ and
$0;(000)$ for Sb$_{2}$Se$_{3}$, indicating a STI and a NI respectively.
One can see that the main difference lies at $\Gamma$ point, i.e.,
$\delta_{1}$ is $-1$ for Bi$_{2}$Se$_{3}$ and $+1$ for Sb$_{2}$Se$_{3}$,
while the other TRIMs share the same parities. We also give the parity
eigenvalues of these 9 bands at $\Gamma$ point, as listed in Table
\ref{Flo:Parity_Gamma}.

\begin{figure}
\includegraphics[width=3.5in]{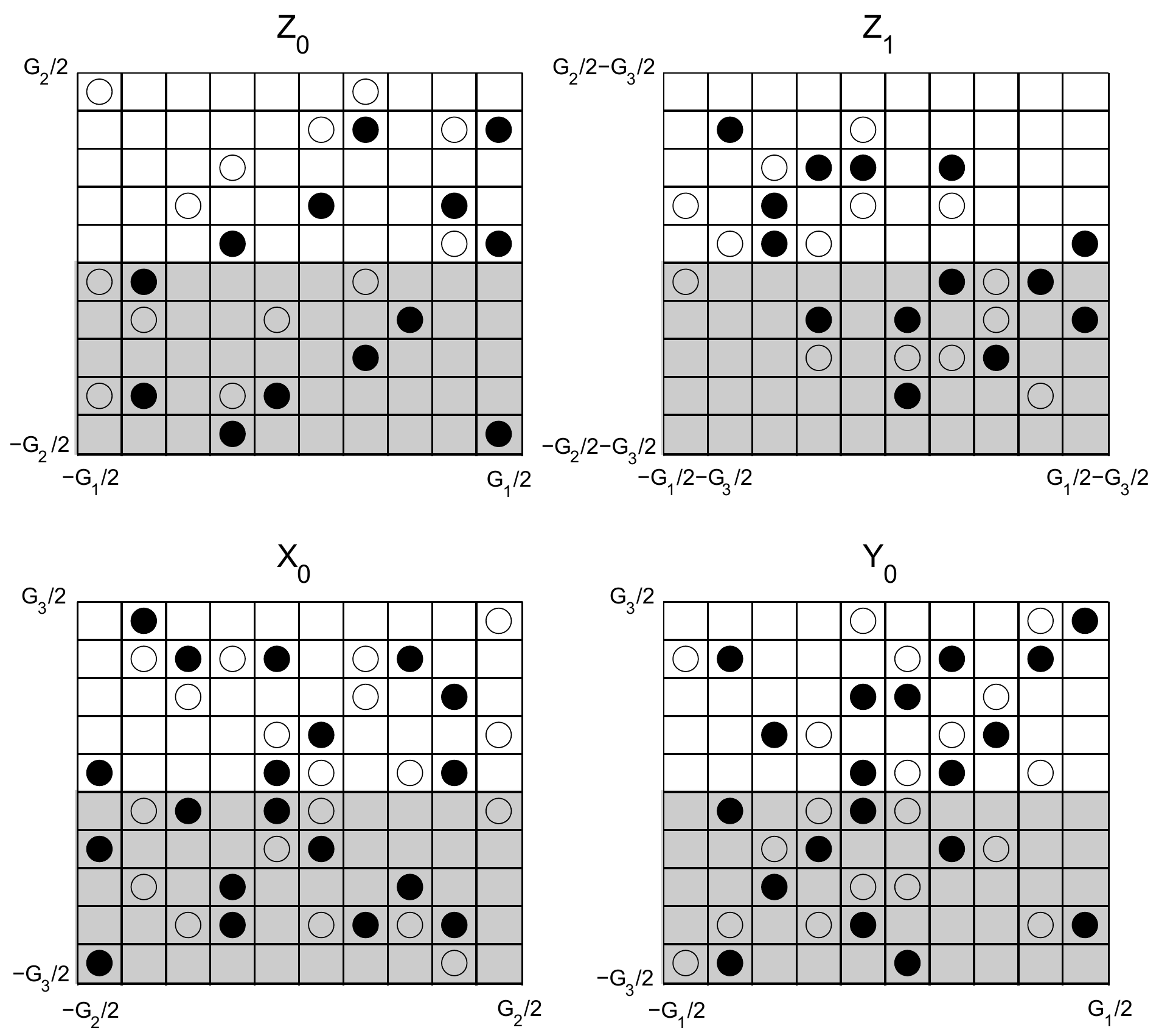}

\caption{The $n$-field configuration for Bi$_{2}$Se$_{3}$ computed under
the time-reversal constraints. The four tori are $T(Z_{0})$, $T(Z_{1})$,
$T(X_{0})$ and $T(Y_{0})$ with the shaded area indicating \emph{half}
of the 2D BZ. The white and black circles denote n = 1 and \textminus{}1,
respectively, while the blank denotes 0. The $Z_{2}$ invariants for
each individual torus is obtained by summing the $n$-field over \emph{half}
of the torus and then moduling 2. These read $z_{0}$ = 1, $z_{1}$
= 0, $x_{0}$ = 1, and $y_{0}$ = 1. The $Z_{2}$ invariants of the
system are 1;(000).}

\label{Flo:Z2_Bi2Se3} 
\end{figure}

\begin{figure}
\includegraphics[width=3.5in]{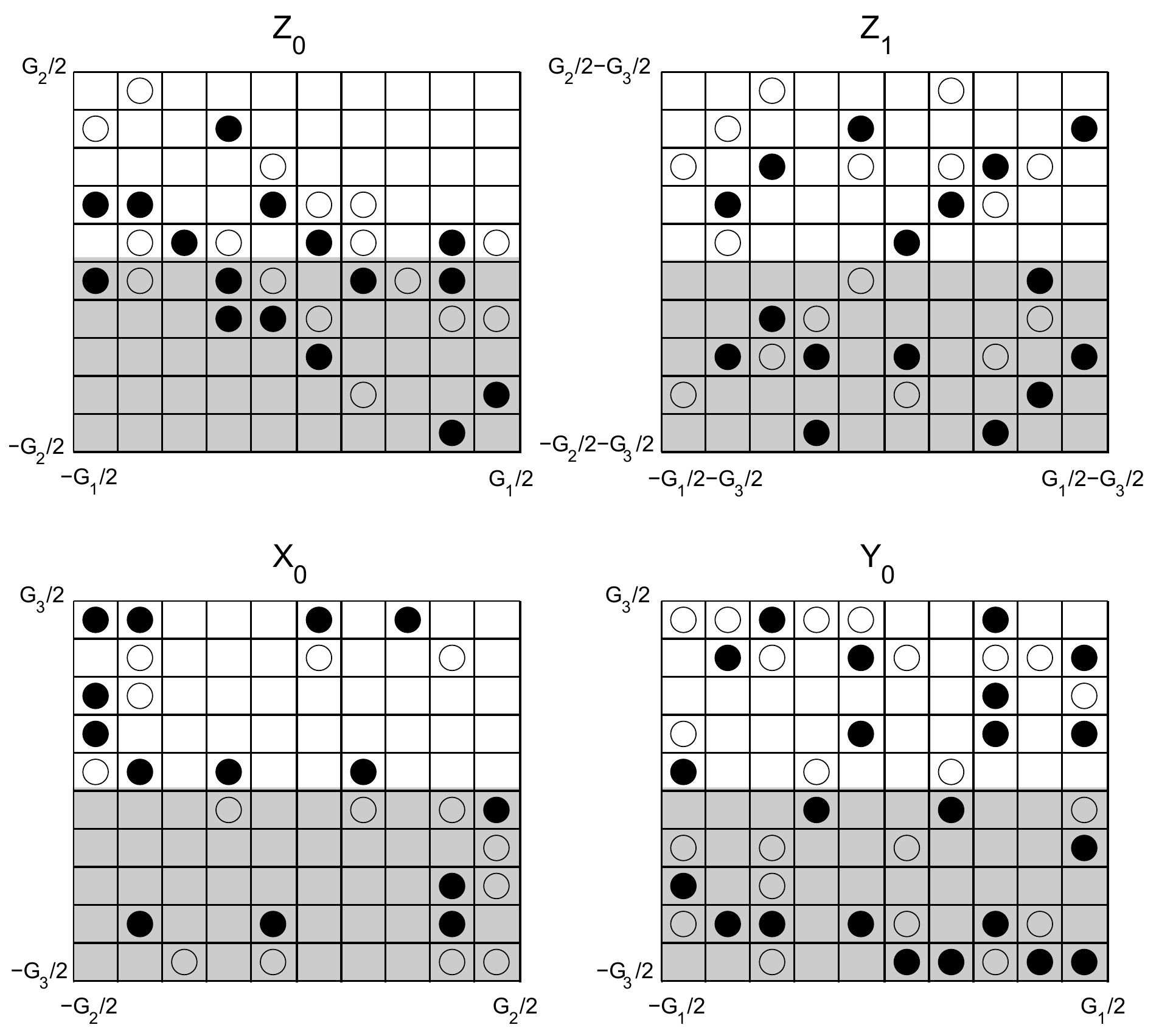}

\caption{The $n$-field configuration of Sb$_{2}$Se$_{3}$. The labels are
the same as Fig. \ref{Flo:Z2_Bi2Se3}. The $Z_{2}$ invariants for
each individual torus read $z_{0}$ = 0, $z_{1}$ = 0, $x_{0}$ =
0, and $y_{0}$ = 0. The $Z_{2}$ invariants of the system are 0;(000).}

\label{Flo:Z2_Sb2Se3} 
\end{figure}

We have also used the lattice calculation of $Z_{2}$ invariants as
a double check. Figure \ref{Flo:Z2_Bi2Se3} shows the $n$-field configuration
for Bi$_{2}$Se$_{3}$. The $Z_{2}$ invariants on each torus are
$z_{0}$ = 1, $z_{1}$ = 0, $x_{0}$ = 1, and $y_{0}$ = 1 by the
sum of the $n$-field in \emph{half} of 2D BZ and then moduling 2.
Total $Z_{2}$ invariants 1;(000) indicate that Bi$_{2}$Se$_{3}$
is a STI. On the other hand, Figure \ref{Flo:Z2_Sb2Se3} shows the
$n$-field configuration for Sb$_{2}$Se$_{3}$ with $z_{0}$ = 0,
$z_{1}$ = 0, $x_{0}$ = 0, and $y_{0}$ = 0 on each torus. Total
$Z_{2}$ invariants 0;(000) indicate that Sb$_{2}$Se$_{3}$ is a
NI. As expected, our lattice calculation of $Z_{2}$ invariants are
the same as parity analysis, and all of these two methods are consistent
with the previous work.\cite{Zhang2009}

\begin{table}
\caption{Parities $\delta_{i}$ at eight TRIMs for Bi$_{2}$Se$_{3}$ and Sb$_{2}$Se$_{3}$.
The relative coordinates in primitive reciprocal-lattice of eight
TRIMs are (0, 0, 0), (0, 0, 0.5), (0, 0.5, 0), (0, 0.5, 0.5), (0.5,
0, 0), (0.5, 0, 0.5), (0.5, 0.5, 0), (0.5, 0.5, 0.5). The $Z_{2}$
invariants are $1;(000)$ for Bi$_{2}$Se$_{3}$ and $0;(000)$ for
Sb$_{2}$Se$_{3}$, which indicate a STI and a NI respectively. }

\label{Flo:Parity_TRIMs}

\begin{ruledtabular}%
\begin{tabular}{cccccccccc}
 & $\delta_{1}$ & $\delta_{2}$  & $\delta_{3}$  & $\delta_{4}$  & $\delta_{5}$  & $\delta_{6}$  & $\delta_{7}$  & $\delta_{8}$  & $\nu_{0};(\nu_{1}\nu_{2}\nu_{3})$\tabularnewline
\hline 
Bi$_{2}$Se$_{3}$  & -1  & +1  & +1  & +1  & +1  & +1  & +1  & +1  & 1;(000)\tabularnewline
Sb$_{2}$Se$_{3}$  & +1  & +1  & +1  & +1  & +1  & +1  & +1  & +1  & 0;(000)\tabularnewline
\end{tabular}\end{ruledtabular} 
\end{table}

\begin{table}
\caption{Parity eigenvalues of Bi$_{2}$Se$_{3}$ and Sb$_{2}$Se$_{3}$ at
$\Gamma$ point for 9 occupied bands. The corresponding band energy
increases from left to right. The parity of $\Gamma$ point, $\delta_{1}$,
is $-1$ for Bi$_{2}$Se$_{3}$ and $+1$ for Sb$_{2}$Se$_{3}$ respectively.}

\label{Flo:Parity_Gamma}

\begin{ruledtabular}%
\begin{tabular}{ccccccccccc}
\multicolumn{1}{c}{} & \multicolumn{9}{c}{} & $\delta_{1}$\tabularnewline
\hline 
Bi$_{2}$Se$_{3}$  & -1  & +1  & +1  & -1  & -1  & +1  & -1  & -1  & +1  & (-1)\tabularnewline
Sb$_{2}$Se$_{3}$  & -1  & -1  & +1  & -1  & +1  & +1  & -1  & -1  & -1  & (+1)\tabularnewline
\end{tabular}\end{ruledtabular} 
\end{table}

\subsection{Noncentrosymmetric systems}

\begin{figure}
\includegraphics[width=3.5in]{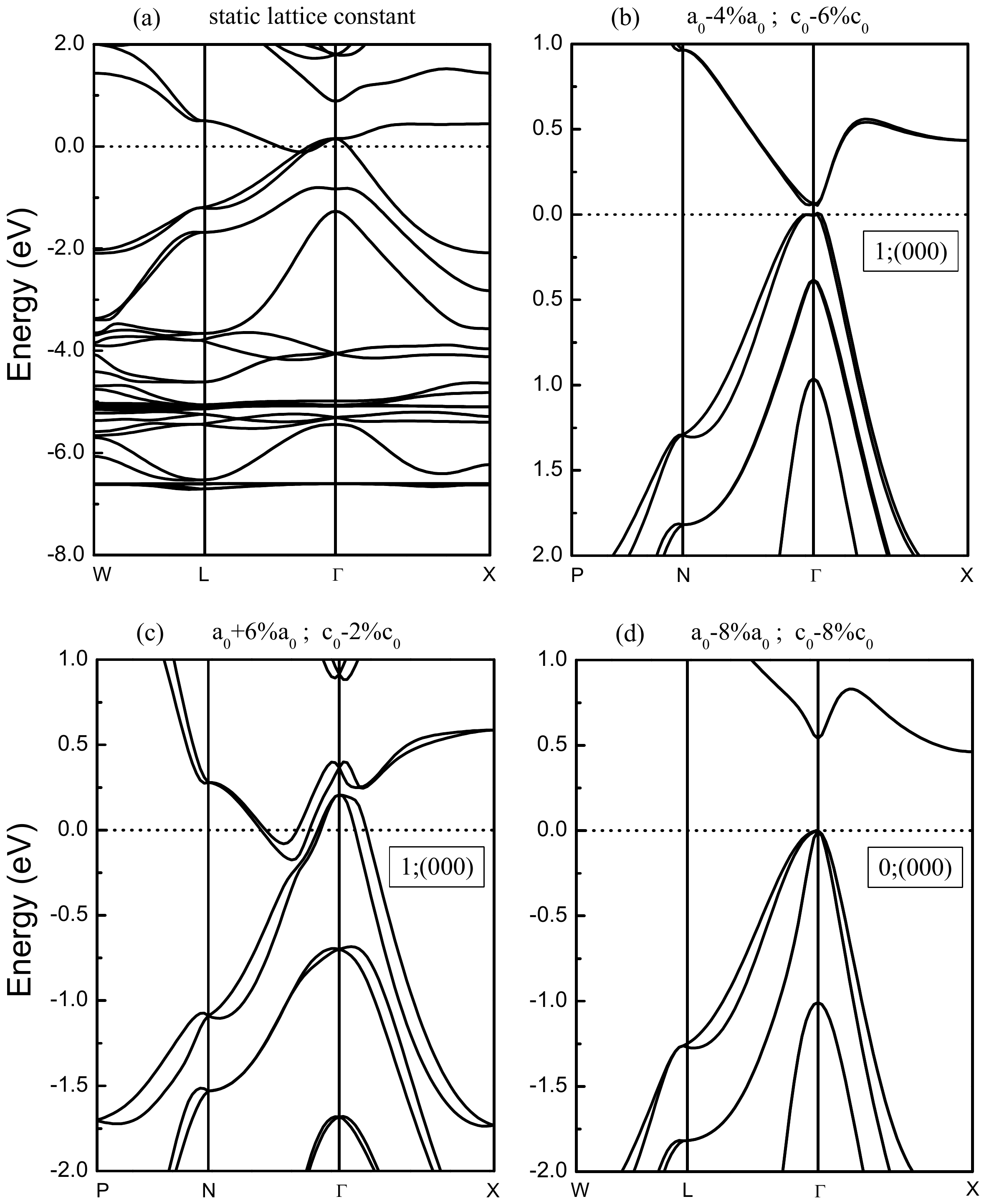}

\caption{Band structures of LuPtBi with the static lattice constant (a) {[}$a_{0}=b_{0}=c_{0}=6.574\text{\AA}${]},
non-hydrostatic strains (b) {[}$a_{0}-4\%a_{0}$, $c_{0}-6\%c_{0}${]}
and (c) {[}$a_{0}+6\%a_{0}$, $c_{0}-2\%c_{0}${]}, and hydrostatic
strain (d) {[}$a_{0}-8\%a_{0}$, $c_{0}-8\%c_{0}${]}. The topological
phases in (a), (b), and (c) are topological insulator, topological
metal, and normal insulator, respectively. The 30 occupied bands (from
$-8$ to about $0$ eV) are used to calculate $Z_{2}$ invariants.}

\label{Flo:Bands_LuPtBi} 
\end{figure}

Having established the effectiveness of our methods in centrosymmetric
systems, we now turn to noncentrosymmetric systems by taking LuPtBi
as the first example. It has been predicted that LuPtBi, as a member
of ternary half-Heusler family, can realize a topological nontrivial
state under uniaxial strain.\cite{Xiao2010a,Chadov2010,Lin2010a,Feng2010,Al-Sawai2010}
The crystal structure of LuPtBi is described by space group $F\bar{4}3m$
(No. 216) with three nonequivalent atoms in a primitive cell. The
calculations were performed using the experimental lattice constant
of 6.574 \AA{}.\cite{Haase2002} As shown in Fig. \ref{Flo:Bands_LuPtBi}(a),
LuPtBi is a semi-metal with small electron and hole pockets around
Fermi level at $\Gamma$ point. The band gap around $\Gamma$ point
can be obtained by applying an uniaxial strain, then 30 occupied bands
(from $-8$ to about $0$ eV) were used to calculate $Z_{2}$ invariants.

As mentioned in our previous works,\cite{Xiao2010a,Feng2010} topological
phases of half-Heusler family are very sensitive to the change of
lattice constants. Generally speaking, hydrostatic expansion leads
to topological nontrivial phases while hydrostatic compression leads
to topological trivial phases. Additionally, one must apply an uniaxial
strain based on hydrostatic strain, i.e., a non-hydrostatic strain,
to realize true topological insulating state because the states around
Fermi level at $\Gamma$ point are fourfold degenerate and protected
by cubic symmetry. Therefore it is necessary to fully understand how
the strain (hydrostatic and non-hydrostatic) acts on the topological
phase in half-Heusler family.

By turning lattice constants $a(=b)$ and $c$, we found three different
topological phases of LuPtBi including STI, TM, and NI, as shown in
Fig. \ref{Flo:Bands_LuPtBi}(b), \ref{Flo:Bands_LuPtBi}(c), and \ref{Flo:Bands_LuPtBi}(d),
respectively. The non-hydrostatic strains can separate the fourfold
degenerate states of valence and conduction bands around $\Gamma$
point. In the case of Fig. \ref{Flo:Bands_LuPtBi}(b), the global
band gap together with $Z_{2}$ invariants $1;(000)$ indicate that
this is a STI. While in the case of Fig. \ref{Flo:Bands_LuPtBi}(c),
it is essentially a metallic state but has local band gap everywhere
in the BZ. The $Z_{2}$ invariants $1;(000)$ show a nontrivial state
which is usually called TM. On the other hand, hydrostatic strain
(large enough compression) can also create a band gap, just like Fig.
\ref{Flo:Bands_LuPtBi}(d), but this is a NI because the $Z_{2}$
invariants are $0;(000)$.

\begin{figure}
\includegraphics[width=3.5in]{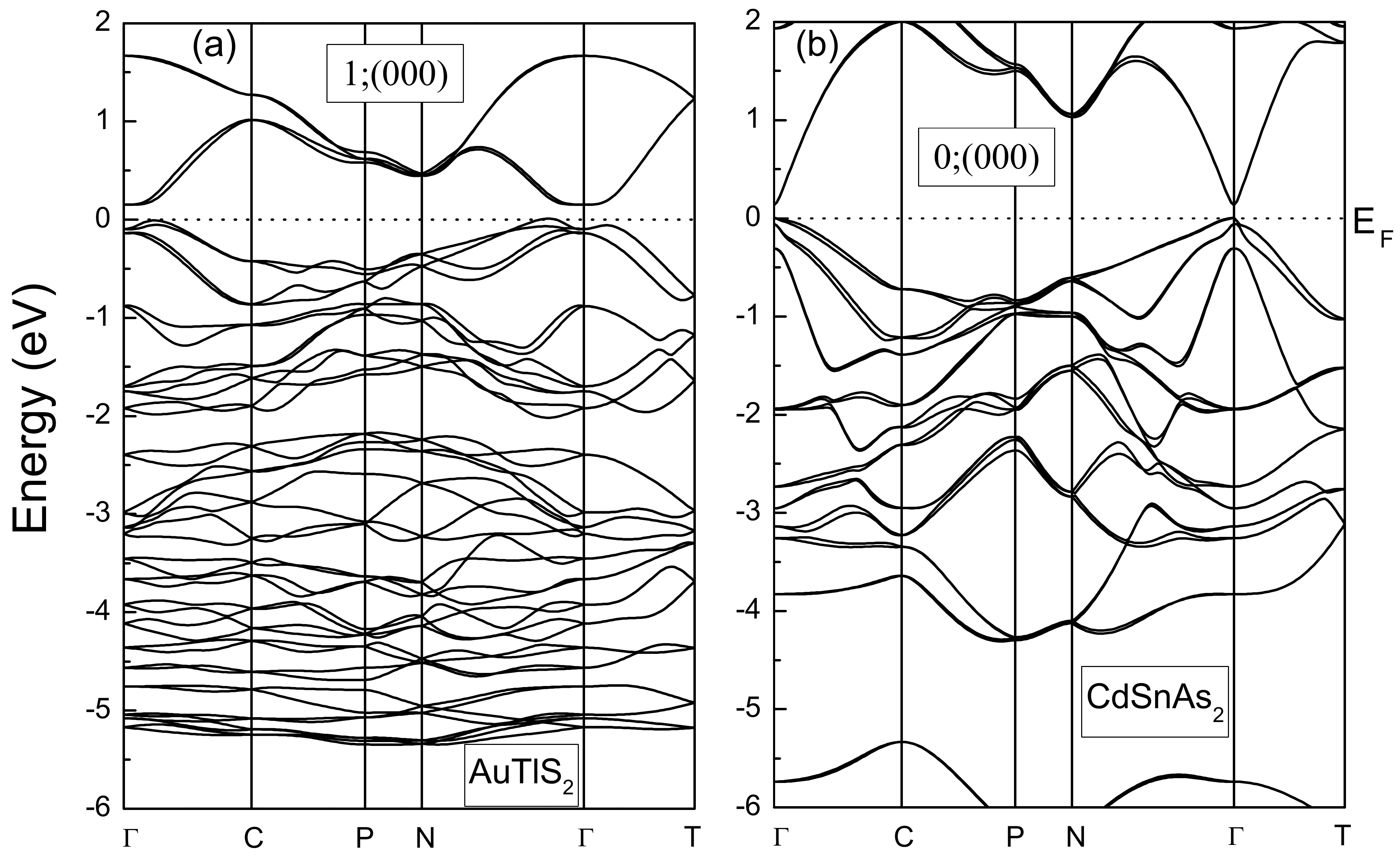}

\caption{Band structures of strong topological insulator AuTlS$_{2}$ with
$Z_{2}$ invariants $1;(000)$ and normal insulator CdSnAs$_{2}$
with $Z_{2}$ invariants $0;(000)$ . The occupied bands which range
from $-6\sim0$ eV are included for calculating $Z_{2}$ invariants,
i.e. 40 bands for AuTlS$_{2}$ and 20 bands for CdSnAs$_{2}$ respectively.
The high-symmetry points in Brillouin zone are the same as Ref.~\onlinecite{Limpijumnong2002}.}

\label{Flo:Bands_chalcopyrite} 
\end{figure}

Ternary chalcopyrite compounds of composition I-III-VI$_{2}$ or II-IV-V$_{2}$
are another important class of noncentrosymmetric TIs. In our previous
work,\cite{Feng2011} we have shown that a large number of ternary
chalcopyrite compounds can realize the topological insulating phase
in their native states. Here we take AuTlS$_{2}$ and CdSnAs$_{2}$
as noncentrosymmetric examples to show our methods for $Z_{2}$ invariants
calculation. The crystal structure of chalcopyrite is described by
the space group $I\bar{4}2d$ (No. 122) with three nonequivalent atoms
in a primitive cell, which can be regarded as a superlattice of the
zinc-blende structure with small structural distortions. The crystal
structure parameters of AuTlS$_{2}$ $\eta=1.016$ and $\delta u=-0.018$
are obtained by first-principles total energy minimization, and the
experimental data $\eta=0.980$ and $\delta u=0.261$\cite{Pamplin1979}
are used for CdSnAs$_{2}$, where $\eta=c/2a$ is the tetragonal distortion
ratio and $\delta u$ is the internal displacement of anion.\cite{Feng2011}
AuTlS$_{2}$ and CdSnAs$_{2}$ are all semiconductors with band gap
of $0.14$ eV and $0.13$ eV, as shown in Fig. \ref{Flo:Bands_chalcopyrite}(a)
and \ref{Flo:Bands_chalcopyrite}(b) respectively. Totally 40 and
20 occupied bands ($-6\sim0$ eV) are used to calculate $Z_{2}$ invariants
for AuTlS$_{2}$ and CdSnAs$_{2}$, respectively. We find that AuTlS$_{2}$
is a STI with the $Z_{2}$ invariants 1;(000) while CdSnAs$_{2}$
is a NI with the $Z_{2}$ invariants 0;(000).

\section{summary}

In summary, we have presented the implementation of first-principles
calculations of topological invariants $Z_{2}$ in both centrosymmetric
and noncentrosymmetric systems within FP-LAPW formalism. Generally,
one can use a lattice version of $Z_{2}$ invariants to identify the
band topology, though in centrosymmetric systems, a simple parity
criterion is possible. The $n$-field configuration depends on a specific
gauge, but the resulting $Z_{2}$ invariants are gauge-invariant.
Our method has two merits: (i) the algorithm implemented in our FP-LAPW
framework is not expensive and the first-principles code can be easily
paralleled; (ii) it is designed as a standard post-process of first-principles
calculations, so the identification of topological nature for a given
material becomes a routine task. Therefore, our method is able to
identify TIs in relatively short time and we anticipate it will speed
up the discovery of new topological insulators in future.
\begin{acknowledgments}
This work is supported by NSF of China (Grants No. 10974231) and the
MOST Project of China (Grant No. 2007CB925000 and 2011CBA00100). W.F.
was supported by the LDRD Program of ORNL. D.X. acknowledges support
by the Materials Sciences and Engineering Division, Office of Basic
Energy Sciences, U.S. Department of Energy. We also acknowledge the
computational resources supported by Texas Advanced Computing Center
(TACC) and Supercomputing Center of Chinese Academy of Sciences (SCCAS). 
\end{acknowledgments}
\appendix

\section{Overlap matrix and its derivatives with the time-reversal operator}

In this appendix, we give the overlap matrix $\left\langle u_{m,\bm{k}}|u_{n,\bm{k+b}}\right\rangle $
and its derivatives with the time-reversal operator $\Theta$, including
$\left\langle u_{m,\bm{k}}|\Theta u_{n,\bm{k+b}}\right\rangle $,
$\left\langle \Theta u_{m,\bm{k}}|u_{n,\bm{k+b}}\right\rangle $,
and $\left\langle \Theta u_{m,\bm{k}}|\Theta u_{n,\bm{k+b}}\right\rangle $,
where $m$ and $n$ stand for band indexes, and $\bm{b}$ stands for
unit vector $\bm{\mu}$ or $\bm{\nu}$ on $\bm{k}$-mesh (see Fig.
\ref{Flo:BZ_2D}).

Firstly, we consider the overlap matrix $\left\langle u_{m,\bm{k}}|u_{n,\bm{k+b}}\right\rangle $
according to Eq. (\ref{eq:Overlap_Matrix}),

\begin{eqnarray}
\left\langle u_{m,\bm{k}}|u_{n,\bm{k+b}}\right\rangle  & = & \left\langle u_{m,\bm{k}}^{\uparrow}|u_{n,\bm{k+b}}^{\uparrow}\right\rangle +\left\langle u_{m,\bm{k}}^{\downarrow}|u_{n,\bm{k+b}}^{\downarrow}\right\rangle .
\end{eqnarray}
 Here we only discuss $\left\langle u_{m,\bm{k}}^{\uparrow}|u_{n,\bm{k+b}}^{\uparrow}\right\rangle $
because that $\left\langle u_{m,\bm{k}}^{\downarrow}|u_{n,\bm{k+b}}^{\downarrow}\right\rangle $
has the similar formulas. Like the BFs, the overlap matrix can also
be divided into two parts: interstitial region and muffin-tin region,

\begin{equation}
\left\langle u_{m,\bm{k}}^{\uparrow}|u_{n,\bm{k+b}}^{\uparrow}\right\rangle =\left\langle u_{m,\bm{k}}^{\uparrow}|u_{n,\bm{k+b}}^{\uparrow}\right\rangle _{I}+{\displaystyle \sum_{\alpha}}\left\langle u_{m,\bm{k}}^{\uparrow}|u_{n,\bm{k+b}}^{\uparrow}\right\rangle _{MT^{\alpha}}.
\end{equation}
 The contribution of interstitial region is

\begin{eqnarray}
\left\langle u_{m,\bm{k}}^{\uparrow}|u_{n,\bm{k+b}}^{\uparrow}\right\rangle _{I} & = & \sum_{{\displaystyle j}}\sum_{{\displaystyle j'}}z_{m,\bm{k},j}^{\uparrow*}z_{n,\bm{k+b},j'}^{\uparrow}\frac{1}{\Omega}\int_{cell}e^{\left[-i\left(\bm{K}_{j}-\bm{K}_{j'}\right)\cdot\bm{r}\right]}\Delta\left(\bm{r}\right)d^{3}r\nonumber \\
 & = & \sum_{{\displaystyle j}}\sum_{{\displaystyle j'}}z_{m,\bm{k},j}^{\uparrow*}z_{n,\bm{k+b},j'}^{\uparrow}\Delta\left(\bm{K}_{j}-\bm{K}_{j'}\right).
\end{eqnarray}
 Here, $\Delta(\bm{r})$ is a step function, it have zero value in
muffin-tin sphere and unit value in interstitial region, and it's
Fourier transform is

\[
\Delta\left(\bm{K}\right)=\delta_{\bm{K},\bm{0}}-\sum_{\alpha}e^{-i\bm{K}\cdot\bm{\tau}^{\alpha}}\frac{4\pi R_{\alpha}^{3}}{\Omega}\frac{j_{1}\left(KR_{\alpha}\right)}{KR_{\alpha}}.
\]
 The contribution of $\alpha$-th muffin-tin sphere is

\begin{eqnarray}
\left\langle u_{m,\bm{k}}^{\uparrow}|u_{n,\bm{k+b}}^{\uparrow}\right\rangle _{MT^{\alpha}} & = & {\displaystyle \int_{MT^{\alpha}}}e^{i\bm{k}\cdot\left(\bm{\tau}^{\alpha}+\bm{r}^{\alpha}\right)}\left[\psi_{m,\bm{k}}^{\uparrow\alpha}(\bm{r})\right]^{*}e^{-i\left(\bm{k+b}\right)\cdot\left(\bm{\tau}^{\alpha}+\bm{r}^{\alpha}\right)}\psi_{n,\bm{k+b}}^{\uparrow\alpha}(\bm{r})d^{3}r\nonumber \\
 & = & {\displaystyle e^{-i\bm{b}\cdot\bm{\tau}^{\alpha}}\int_{MT^{\alpha}}}\left[\psi_{m,\bm{k}}^{\uparrow\alpha}(\bm{r})\right]^{*}\psi_{n,\bm{k+b}}^{\uparrow\alpha}(\bm{r})e^{-i\bm{b}\cdot\bm{r}^{\alpha}}d^{3}r.
\end{eqnarray}
 Using the Rayleigh plane-wave expansion

\begin{equation}
e^{-i\bm{b}\cdot\bm{r}^{\alpha}}=4\pi{\displaystyle \sum_{l^{\prime\prime}m^{\prime\prime}}}\left(-i\right)^{l^{\prime\prime}}Y_{l^{\prime\prime}m^{\prime\prime}}^{*}\left(\hat{\bm{b}}\right)Y_{l^{\prime\prime}m^{\prime\prime}}\left(\hat{\bm{r}}^{\alpha}\right)j_{l^{\prime\prime}}\left(br^{\alpha}\right),
\end{equation}
 where $r^{\alpha}=\left|\bm{r}^{\alpha}\right|$, $b=\left|\bm{b}\right|$,
and $j_{l^{\prime\prime}}\left(br^{\alpha}\right)$ is the spherical
bessel function. Then,

\begin{align}
\left\langle u_{m,\bm{k}}^{\uparrow}|u_{n,\bm{k+b}}^{\uparrow}\right\rangle _{MT^{\alpha}} & =4\pi e^{-i\bm{b}\cdot\bm{\tau}^{\alpha}}{\displaystyle \sum_{l^{\prime\prime}m^{\prime\prime}}}\left(-i\right)^{l^{\prime\prime}}Y_{l^{\prime\prime}m^{\prime\prime}}^{*}\left(\hat{\bm{b}}\right)\nonumber \\
 & \times{\displaystyle \sum_{lm}\sum_{l^{\prime}m^{\prime}}}\left\{ \left[A_{lm}^{\uparrow\alpha}\left(m,\bm{k}\right)\right]^{*}A_{l^{\prime}m^{\prime}}^{\uparrow\alpha}\left(n,\bm{k+b}\right)\left[u_{l,1}^{\uparrow,\alpha}u_{l^{\prime},1}^{\uparrow,\alpha}j_{l^{\prime\prime},b}\right]\right.\nonumber \\
 & +\left[A_{lm}^{\uparrow\alpha}\left(m,\bm{k}\right)\right]^{*}B_{l^{\prime}m^{\prime}}^{\uparrow\alpha}\left(n,\bm{k+b}\right)\left[u_{l,1}^{\uparrow,\alpha}\dot{u}_{l^{\prime},1}^{\uparrow,\alpha}j_{l^{\prime\prime},b}\right]\nonumber \\
 & +\left[A_{lm}^{\uparrow\alpha}\left(m,\bm{k}\right)\right]^{*}C_{l^{\prime}m^{\prime}}^{\uparrow\alpha}\left(n,\bm{k+b}\right)\left[u_{l,1}^{\uparrow,\alpha}u_{l^{\prime},2}^{\uparrow,\alpha}j_{l^{\prime\prime},b}\right]\nonumber \\
 & +\left[A_{lm}^{\uparrow\alpha}\left(m,\bm{k}\right)\right]^{*}D_{l^{\prime}m^{\prime}}^{\uparrow\alpha}\left(n,\bm{k+b}\right)\left[u_{l,1}^{\uparrow,\alpha}u_{l^{\prime},1/2}^{\uparrow,\alpha}j_{l^{\prime\prime},b}\right]\nonumber \\
 & +\left[B_{lm}^{\uparrow\alpha}\left(m,\bm{k}\right)\right]^{*}A_{l^{\prime}m^{\prime}}^{\uparrow\alpha}\left(n,\bm{k+b}\right)\left[\dot{u}_{l,1}^{\uparrow,\alpha}u_{l^{\prime},1}^{\uparrow,\alpha}j_{l^{\prime\prime},b}\right]\nonumber \\
 & +\left[B_{lm}^{\uparrow\alpha}\left(m,\bm{k}\right)\right]^{*}B_{l^{\prime}m^{\prime}}^{\uparrow\alpha}\left(n,\bm{k+b}\right)\left[\dot{u}_{l,1}^{\uparrow,\alpha}\dot{u}_{l^{\prime},1}^{\uparrow,\alpha}j_{l^{\prime\prime},b}\right]\nonumber \\
 & +\left[B_{lm}^{\uparrow\alpha}\left(m,\bm{k}\right)\right]^{*}C_{l^{\prime}m^{\prime}}^{\uparrow\alpha}\left(n,\bm{k+b}\right)\left[\dot{u}_{l,1}^{\uparrow,\alpha}u_{l^{\prime},2}^{\uparrow,\alpha}j_{l^{\prime\prime},b}\right]\nonumber \\
 & +\left[B_{lm}^{\uparrow\alpha}\left(m,\bm{k}\right)\right]^{*}D_{l^{\prime}m^{\prime}}^{\uparrow\alpha}\left(n,\bm{k+b}\right)\left[\dot{u}_{l,1}^{\uparrow,\alpha}u_{l^{\prime},1/2}^{\uparrow,\alpha}j_{l^{\prime\prime},b}\right]\nonumber \\
 & +\left[C_{lm}^{\uparrow\alpha}\left(m,\bm{k}\right)\right]^{*}A_{l^{\prime}m^{\prime}}^{\uparrow\alpha}\left(n,\bm{k+b}\right)\left[u_{l,2}^{\uparrow,\alpha}u_{l^{\prime},1}^{\uparrow,\alpha}j_{l^{\prime\prime},b}\right]\nonumber \\
 & +\left[C_{lm}^{\uparrow\alpha}\left(m,\bm{k}\right)\right]^{*}B_{l^{\prime}m^{\prime}}^{\uparrow\alpha}\left(n,\bm{k+b}\right)\left[u_{l,2}^{\uparrow,\alpha}\dot{u}_{l^{\prime},1}^{\uparrow,\alpha}j_{l^{\prime\prime},b}\right]\nonumber \\
 & +\left[C_{lm}^{\uparrow\alpha}\left(m,\bm{k}\right)\right]^{*}C_{l^{\prime}m^{\prime}}^{\uparrow\alpha}\left(n,\bm{k+b}\right)\left[u_{l,2}^{\uparrow,\alpha}u_{l^{\prime},2}^{\uparrow,\alpha}j_{l^{\prime\prime},b}\right]\nonumber \\
 & +\left[C_{lm}^{\uparrow\alpha}\left(m,\bm{k}\right)\right]^{*}D_{l^{\prime}m^{\prime}}^{\uparrow\alpha}\left(n,\bm{k+b}\right)\left[u_{l,2}^{\uparrow,\alpha}u_{l^{\prime},1/2}^{\uparrow,\alpha}j_{l^{\prime\prime},b}\right]\nonumber \\
 & +\left[D_{lm}^{\uparrow\alpha}\left(m,\bm{k}\right)\right]^{*}A_{l^{\prime}m^{\prime}}^{\uparrow\alpha}\left(n,\bm{k+b}\right)\left[u_{l,1/2}^{\uparrow,\alpha}u_{l^{\prime},1}^{\uparrow,\alpha}j_{l^{\prime\prime},b}\right]\nonumber \\
 & +\left[D_{lm}^{\uparrow\alpha}\left(m,\bm{k}\right)\right]^{*}B_{l^{\prime}m^{\prime}}^{\uparrow\alpha}\left(n,\bm{k+b}\right)\left[u_{l,1/2}^{\uparrow,\alpha}\dot{u}_{l^{\prime},1}^{\uparrow,\alpha}j_{l^{\prime\prime},b}\right]\nonumber \\
 & +\left[D_{lm}^{\uparrow\alpha}\left(m,\bm{k}\right)\right]^{*}C_{l^{\prime}m^{\prime}}^{\uparrow\alpha}\left(n,\bm{k+b}\right)\left[u_{l,1/2}^{\uparrow,\alpha}u_{l^{\prime},2}^{\uparrow,\alpha}j_{l^{\prime\prime},b}\right]\nonumber \\
 & \left.+\left[D_{lm}^{\uparrow\alpha}\left(m,\bm{k}\right)\right]^{*}D_{l^{\prime}m^{\prime}}^{\uparrow\alpha}\left(n,\bm{k+b}\right)\left[u_{l,1/2}^{\uparrow,\alpha}u_{l^{\prime},1/2}^{\uparrow,\alpha}j_{l^{\prime\prime},b}\right]\right\} G_{ll^{\prime}l^{\prime\prime}}^{mm^{\prime}m^{\prime\prime}}.
\end{align}
 Therefore, the matrix elements $\left\langle u_{m,\bm{k}}^{\uparrow}|u_{n,\bm{k+b}}^{\uparrow}\right\rangle _{MT^{\alpha}}$
are constructed by two parts: radial integrals and angular integrals.
The radial integrals are 
\begin{align}
\left[u_{l,1}^{\uparrow,\alpha}u_{l^{\prime},1}^{\uparrow,\alpha}j_{l^{\prime\prime},b}\right] & =\int_{0}^{R^{\alpha}}r^{2}u_{l}^{\uparrow}\left(r^{\alpha},E_{l,1}^{\alpha}\right)u_{l^{\prime}}^{\uparrow}\left(r^{\alpha},E_{l^{\prime},1}^{\alpha}\right)j_{l^{\prime\prime}}\left(br^{\alpha}\right)dr,\nonumber \\
\left[u_{l,1}^{\uparrow,\alpha}\dot{u}_{l^{\prime},1}^{\uparrow,\alpha}j_{l^{\prime\prime},b}\right] & =\int_{0}^{R^{\alpha}}r^{2}u_{l}^{\uparrow}\left(r^{\alpha},E_{l,1}^{\alpha}\right)\dot{u}_{l^{\prime}}^{\uparrow}\left(r^{\alpha},E_{l^{\prime},1}^{\alpha}\right)j_{l^{\prime\prime}}\left(br^{\alpha}\right)dr,\nonumber \\
\left[u_{l,1}^{\uparrow,\alpha}u_{l^{\prime},2}^{\uparrow,\alpha}j_{l^{\prime\prime},b}\right] & =\int_{0}^{R^{\alpha}}r^{2}u_{l}^{\uparrow}\left(r^{\alpha},E_{l,1}^{\alpha}\right)u_{l^{\prime}}^{\uparrow}\left(r^{\alpha},E_{l^{\prime},2}^{\alpha}\right)j_{l^{\prime\prime}}\left(br^{\alpha}\right)dr,\nonumber \\
\left[u_{l,1}^{\uparrow,\alpha}u_{l^{\prime},1/2}^{\uparrow,\alpha}j_{l^{\prime\prime},b}\right] & =\int_{0}^{R^{\alpha}}r^{2}u_{l}^{\uparrow}\left(r^{\alpha},E_{l,1}^{\alpha}\right)u_{l^{\prime}}^{\uparrow}\left(r^{\alpha},E_{l^{\prime},1/2}^{\alpha}\right)j_{l^{\prime\prime}}\left(br^{\alpha}\right)dr,\nonumber \\
\left[\dot{u}_{l,1}^{\uparrow,\alpha}u_{l^{\prime},1}^{\uparrow,\alpha}j_{l^{\prime\prime},b}\right] & =\int_{0}^{R^{\alpha}}r^{2}\dot{u}_{l}^{\uparrow}\left(r^{\alpha},E_{l,1}^{\alpha}\right)u_{l^{\prime}}^{\uparrow}\left(r^{\alpha},E_{l^{\prime},1}^{\alpha}\right)j_{l^{\prime\prime}}\left(br^{\alpha}\right)dr,\nonumber \\
\left[\dot{u}_{l,1}^{\uparrow,\alpha}\dot{u}_{l^{\prime},1}^{\uparrow,\alpha}j_{l^{\prime\prime},b}\right] & =\int_{0}^{R^{\alpha}}r^{2}\dot{u}_{l}^{\uparrow}\left(r^{\alpha},E_{l,1}^{\alpha}\right)\dot{u}_{l^{\prime}}^{\uparrow}\left(r^{\alpha},E_{l^{\prime},1}^{\alpha}\right)j_{l^{\prime\prime}}\left(br^{\alpha}\right)dr,\nonumber \\
\left[\dot{u}_{l,1}^{\uparrow,\alpha}u_{l^{\prime},2}^{\uparrow,\alpha}j_{l^{\prime\prime},b}\right] & =\int_{0}^{R^{\alpha}}r^{2}\dot{u}_{l}^{\uparrow}\left(r^{\alpha},E_{l,1}^{\alpha}\right)u_{l^{\prime}}^{\uparrow}\left(r^{\alpha},E_{l^{\prime},2}^{\alpha}\right)j_{l^{\prime\prime}}\left(br^{\alpha}\right)dr,\nonumber \\
\left[\dot{u}_{l,1}^{\uparrow,\alpha}u_{l^{\prime},1/2}^{\uparrow,\alpha}j_{l^{\prime\prime},b}\right] & =\int_{0}^{R^{\alpha}}r^{2}\dot{u}_{l}^{\uparrow}\left(r^{\alpha},E_{l,1}^{\alpha}\right)u_{l^{\prime}}^{\uparrow}\left(r^{\alpha},E_{l^{\prime},1/2}^{\alpha}\right)j_{l^{\prime\prime}}\left(br^{\alpha}\right)dr,\nonumber \\
\left[u_{l,2}^{\uparrow,\alpha}u_{l^{\prime},1}^{\uparrow,\alpha}j_{l^{\prime\prime},b}\right] & =\int_{0}^{R^{\alpha}}r^{2}u_{l}^{\uparrow}\left(r^{\alpha},E_{l,2}^{\alpha}\right)u_{l^{\prime}}^{\uparrow}\left(r^{\alpha},E_{l^{\prime},1}^{\alpha}\right)j_{l^{\prime\prime}}\left(br^{\alpha}\right)dr,\nonumber \\
\left[u_{l,2}^{\uparrow,\alpha}\dot{u}_{l^{\prime},1}^{\uparrow,\alpha}j_{l^{\prime\prime},b}\right] & =\int_{0}^{R^{\alpha}}r^{2}u_{l}^{\uparrow}\left(r^{\alpha},E_{l,2}^{\alpha}\right)\dot{u}_{l^{\prime}}^{\uparrow}\left(r^{\alpha},E_{l^{\prime},1}^{\alpha}\right)j_{l^{\prime\prime}}\left(br^{\alpha}\right)dr,\nonumber \\
\left[u_{l,2}^{\uparrow,\alpha}u_{l^{\prime},2}^{\uparrow,\alpha}j_{l^{\prime\prime},b}\right] & =\int_{0}^{R^{\alpha}}r^{2}u_{l}^{\uparrow}\left(r^{\alpha},E_{l,2}^{\alpha}\right)u_{l^{\prime}}^{\uparrow}\left(r^{\alpha},E_{l^{\prime},2}^{\alpha}\right)j_{l^{\prime\prime}}\left(br^{\alpha}\right)dr,\nonumber \\
\left[u_{l,2}^{\uparrow,\alpha}u_{l^{\prime},1/2}^{\uparrow,\alpha}j_{l^{\prime\prime},b}\right] & =\int_{0}^{R^{\alpha}}r^{2}u_{l}^{\uparrow}\left(r^{\alpha},E_{l,2}^{\alpha}\right)u_{l^{\prime}}^{\uparrow}\left(r^{\alpha},E_{l^{\prime},1/2}^{\alpha}\right)j_{l^{\prime\prime}}\left(br^{\alpha}\right)dr,\nonumber \\
\left[u_{l,1/2}^{\uparrow,\alpha}u_{l^{\prime},1}^{\uparrow,\alpha}j_{l^{\prime\prime},b}\right] & =\int_{0}^{R^{\alpha}}r^{2}u_{l}^{\uparrow}\left(r^{\alpha},E_{l,1/2}^{\alpha}\right)u_{l^{\prime}}^{\uparrow}\left(r^{\alpha},E_{l^{\prime},1}^{\alpha}\right)j_{l^{\prime\prime}}\left(br^{\alpha}\right)dr,\nonumber \\
\left[u_{l,1/2}^{\uparrow,\alpha}\dot{u}_{l^{\prime},1}^{\uparrow,\alpha}j_{l^{\prime\prime},b}\right] & =\int_{0}^{R^{\alpha}}r^{2}u_{l}^{\uparrow}\left(r^{\alpha},E_{l,1/2}^{\alpha}\right)\dot{u}_{l^{\prime}}^{\uparrow}\left(r^{\alpha},E_{l^{\prime},1}^{\alpha}\right)j_{l^{\prime\prime}}\left(br^{\alpha}\right)dr,\nonumber \\
\left[u_{l,1/2}^{\uparrow,\alpha}u_{l^{\prime},2}^{\uparrow,\alpha}j_{l^{\prime\prime},b}\right] & =\int_{0}^{R^{\alpha}}r^{2}u_{l}^{\uparrow}\left(r^{\alpha},E_{l,1/2}^{\alpha}\right)u_{l^{\prime}}^{\uparrow}\left(r^{\alpha},E_{l^{\prime},2}^{\alpha}\right)j_{l^{\prime\prime}}\left(br^{\alpha}\right)dr,\nonumber \\
\left[u_{l,1/2}^{\uparrow,\alpha}u_{l^{\prime},1/2}^{\uparrow,\alpha}j_{l^{\prime\prime},b}\right] & =\int_{0}^{R^{\alpha}}r^{2}u_{l}^{\uparrow}\left(r^{\alpha},E_{l,1/2}^{\alpha}\right)u_{l^{\prime}}^{\uparrow}\left(r^{\alpha},E_{l^{\prime},1/2}^{\alpha}\right)j_{l^{\prime\prime}}\left(br^{\alpha}\right)dr,
\end{align}
 and the angular integrals is the Gaunt coefficients

\begin{equation}
G_{ll^{\prime}l^{\prime\prime}}^{mm^{\prime}m^{\prime\prime}}=\int Y_{lm}^{*}\left(\hat{\bm{r}}\right)Y_{l^{\prime}m^{\prime}}\left(\hat{\bm{r}}\right)Y_{l^{\prime\prime}m^{\prime\prime}}\left(\hat{\bm{r}}\right)d\bm{\Omega}.
\end{equation}

Secondly, we consider the matrix element $\left\langle u_{m,\bm{k}}|\Theta u_{n,\bm{k+b}}\right\rangle $,

\begin{eqnarray}
\left\langle u_{m,\bm{k}}|\Theta u_{n,\bm{k+b}}\right\rangle  & = & -\left\langle u_{m,\bm{k}}^{\uparrow}|u_{n,\bm{k+b}}^{\downarrow*}\right\rangle +\left\langle u_{m,\bm{k}}^{\downarrow}|u_{n,\bm{k+b}}^{\uparrow*}\right\rangle ,
\end{eqnarray}
 and take $\left\langle u_{m,\bm{k}}^{\uparrow}|u_{n,\bm{k+b}}^{\downarrow*}\right\rangle $
as example, it can be divided into two parts:

\begin{equation}
\left\langle u_{m,\bm{k}}^{\uparrow}|u_{n,\bm{k+b}}^{\downarrow*}\right\rangle =\left\langle u_{m,\bm{k}}^{\uparrow}|u_{n,\bm{k+b}}^{\downarrow*}\right\rangle _{I}+{\displaystyle \sum_{\alpha}}\left\langle u_{m,\bm{k}}^{\uparrow}|u_{n,\bm{k+b}}^{\downarrow*}\right\rangle _{MT^{\alpha}}.
\end{equation}
 Within the interstitial region,

\begin{eqnarray}
\left\langle u_{m,\bm{k}}^{\uparrow}|u_{n,\bm{k+b}}^{\downarrow*}\right\rangle _{I} & = & \sum_{{\displaystyle j}}\sum_{{\displaystyle j'}}z_{m\bm{k},j}^{\uparrow*}z_{n\bm{k+b},j'}^{\downarrow*}\frac{1}{\Omega}\int_{cell}e^{\left[-i\left(\bm{K}_{j}+\bm{K}_{j'}\right)\cdot\bm{r}\right]}\Theta\left(\bm{r}\right)d^{3}r\nonumber \\
 & = & \sum_{{\displaystyle j}}\sum_{{\displaystyle j'}}z_{m\bm{k},j}^{\uparrow*}z_{n\bm{k+b},j'}^{\downarrow*}\Theta\left(\bm{K}_{j}+\bm{K}_{j'}\right),
\end{eqnarray}
 while inside the muffin-tin region ($\alpha$-th atom sphere),

\begin{align}
\left\langle u_{m,\bm{k}}^{\uparrow}|u_{n,\bm{k+b}}^{\downarrow*}\right\rangle _{MT^{\alpha}} & =4\pi e^{i\left(2\bm{k+b}\right)\cdot\bm{\tau}^{\alpha}}{\displaystyle \sum_{l^{\prime\prime}m^{\prime\prime}}}i^{l^{\prime\prime}}Y_{l^{\prime\prime}m^{\prime\prime}}^{*}\left(\widehat{2\bm{k+b}}\right)\nonumber \\
 & \times{\displaystyle \sum_{lm}\sum_{l^{\prime}m^{\prime}}}\left\{ \left[A_{lm}^{\uparrow\alpha}\left(m,\bm{k}\right)\right]^{*}\left[A_{l^{\prime}m^{\prime}}^{\downarrow\alpha}\left(n,\bm{k+b}\right)\right]^{*}\left[u_{l,1}^{\uparrow,\alpha}u_{l^{\prime},1}^{\downarrow,\alpha}j_{l^{\prime\prime},2k+b}\right]\right.\nonumber \\
 & +\left[A_{lm}^{\uparrow\alpha}\left(m,\bm{k}\right)\right]^{*}\left[B_{l^{\prime}m^{\prime}}^{\downarrow\alpha}\left(n,\bm{k+b}\right)\right]^{*}\left[u_{l,1}^{\uparrow,\alpha}\dot{u}_{l^{\prime},1}^{\downarrow,\alpha}j_{l^{\prime\prime},2k+b}\right]\nonumber \\
 & +\left[A_{lm}^{\uparrow\alpha}\left(m,\bm{k}\right)\right]^{*}\left[C_{l^{\prime}m^{\prime}}^{\downarrow\alpha}\left(n,\bm{k+b}\right)\right]^{*}\left[u_{l,1}^{\uparrow,\alpha}u_{l^{\prime},2}^{\downarrow,\alpha}j_{l^{\prime\prime},2k+b}\right]\nonumber \\
 & +\left[A_{lm}^{\uparrow\alpha}\left(m,\bm{k}\right)\right]^{*}\left[D_{l^{\prime}m^{\prime}}^{\downarrow\alpha}\left(n,\bm{k+b}\right)\right]^{*}\left[u_{l,1}^{\uparrow,\alpha}u_{l^{\prime},1/2}^{\downarrow,\alpha}j_{l^{\prime\prime},2k+b}\right]\nonumber \\
 & +\left[B_{lm}^{\uparrow\alpha}\left(m,\bm{k}\right)\right]^{*}\left[A_{l^{\prime}m^{\prime}}^{\downarrow\alpha}\left(n,\bm{k+b}\right)\right]^{*}\left[\dot{u}_{l,1}^{\uparrow,\alpha}u_{l^{\prime},1}^{\downarrow,\alpha}j_{l^{\prime\prime},2k+b}\right]\nonumber \\
 & +\left[B_{lm}^{\uparrow\alpha}\left(m,\bm{k}\right)\right]^{*}\left[B_{l^{\prime}m^{\prime}}^{\downarrow\alpha}\left(n,\bm{k+b}\right)\right]^{*}\left[\dot{u}_{l,1}^{\uparrow,\alpha}\dot{u}_{l^{\prime},1}^{\downarrow,\alpha}j_{l^{\prime\prime},2k+b}\right]\nonumber \\
 & +\left[B_{lm}^{\uparrow\alpha}\left(m,\bm{k}\right)\right]^{*}\left[C_{l^{\prime}m^{\prime}}^{\downarrow\alpha}\left(n,\bm{k+b}\right)\right]^{*}\left[\dot{u}_{l,1}^{\uparrow,\alpha}u_{l^{\prime},2}^{\downarrow,\alpha}j_{l^{\prime\prime},2k+b}\right]\nonumber \\
 & +\left[B_{lm}^{\uparrow\alpha}\left(m,\bm{k}\right)\right]^{*}\left[D_{l^{\prime}m^{\prime}}^{\downarrow\alpha}\left(n,\bm{k+b}\right)\right]^{*}\left[\dot{u}_{l,1}^{\uparrow,\alpha}u_{l^{\prime},1/2}^{\downarrow,\alpha}j_{l^{\prime\prime},2k+b}\right]\nonumber \\
 & +\left[C_{lm}^{\uparrow\alpha}\left(m,\bm{k}\right)\right]^{*}\left[A_{l^{\prime}m^{\prime}}^{\downarrow\alpha}\left(n,\bm{k+b}\right)\right]^{*}\left[u_{l,2}^{\uparrow,\alpha}u_{l^{\prime},1}^{\downarrow,\alpha}j_{l^{\prime\prime},2k+b}\right]\nonumber \\
 & +\left[C_{lm}^{\uparrow\alpha}\left(m,\bm{k}\right)\right]^{*}\left[B_{l^{\prime}m^{\prime}}^{\downarrow\alpha}\left(n,\bm{k+b}\right)\right]^{*}\left[u_{l,2}^{\uparrow,\alpha}\dot{u}_{l^{\prime},1}^{\downarrow,\alpha}j_{l^{\prime\prime},2k+b}\right]\nonumber \\
 & +\left[C_{lm}^{\uparrow\alpha}\left(m,\bm{k}\right)\right]^{*}\left[C_{l^{\prime}m^{\prime}}^{\downarrow\alpha}\left(n,\bm{k+b}\right)\right]^{*}\left[u_{l,2}^{\uparrow,\alpha}u_{l^{\prime},2}^{\downarrow,\alpha}j_{l^{\prime\prime},2k+b}\right]\nonumber \\
 & +\left[C_{lm}^{\uparrow\alpha}\left(m,\bm{k}\right)\right]^{*}\left[D_{l^{\prime}m^{\prime}}^{\downarrow\alpha}\left(n,\bm{k+b}\right)\right]^{*}\left[u_{l,2}^{\uparrow,\alpha}u_{l^{\prime},1/2}^{\downarrow,\alpha}j_{l^{\prime\prime},2k+b}\right]\nonumber \\
 & +\left[D_{lm}^{\uparrow\alpha}\left(m,\bm{k}\right)\right]^{*}\left[A_{l^{\prime}m^{\prime}}^{\downarrow\alpha}\left(n,\bm{k+b}\right)\right]^{*}\left[u_{l,1/2}^{\uparrow,\alpha}u_{l^{\prime},1}^{\downarrow,\alpha}j_{l^{\prime\prime},2k+b}\right]\nonumber \\
 & +\left[D_{lm}^{\uparrow\alpha}\left(m,\bm{k}\right)\right]^{*}\left[B_{l^{\prime}m^{\prime}}^{\downarrow\alpha}\left(n,\bm{k+b}\right)\right]^{*}\left[u_{l,1/2}^{\uparrow,\alpha}\dot{u}_{l^{\prime},1}^{\downarrow,\alpha}j_{l^{\prime\prime},2k+b}\right]\nonumber \\
 & +\left[D_{lm}^{\uparrow\alpha}\left(m,\bm{k}\right)\right]^{*}\left[C_{l^{\prime}m^{\prime}}^{\downarrow\alpha}\left(n,\bm{k+b}\right)\right]^{*}\left[u_{l,1/2}^{\uparrow,\alpha}u_{l^{\prime},2}^{\downarrow,\alpha}j_{l^{\prime\prime},2k+b}\right]\nonumber \\
 & \left.+\left[D_{lm}^{\uparrow\alpha}\left(m,\bm{k}\right)\right]^{*}\left[D_{l^{\prime}m^{\prime}}^{\downarrow\alpha}\left(n,\bm{k+b}\right)\right]^{*}\left[u_{l,1/2}^{\uparrow,\alpha}u_{l^{\prime},1/2}^{\downarrow,\alpha}j_{l^{\prime\prime},2k+b}\right]\right\} \left(-1\right)^{m^{\prime}}G_{ll^{\prime}l^{\prime\prime}}^{m-m^{\prime}m^{\prime\prime}}.
\end{align}

Thirdly, we consider the matrix element $\left\langle \Theta u_{m,\bm{k}}|u_{n,\bm{k+b}}\right\rangle $,

\begin{eqnarray}
\left\langle \Theta u_{m,\bm{k}}|u_{n,\bm{k+b}}\right\rangle  & = & -\left\langle u_{m,\bm{k}}^{\downarrow*}|u_{n,\bm{k+b}}^{\uparrow}\right\rangle +\left\langle u_{m,\bm{k}}^{\uparrow*}|u_{n,\bm{k+b}}^{\downarrow}\right\rangle ,
\end{eqnarray}
 and take $\left\langle u_{m,\bm{k}}^{\downarrow*}|u_{n,\bm{k+b}}^{\uparrow}\right\rangle $
as example, it can be divided into two parts

\begin{equation}
\left\langle u_{m,\bm{k}}^{\downarrow*}|u_{n,\bm{k+b}}^{\uparrow}\right\rangle =\left\langle u_{m,\bm{k}}^{\downarrow*}|u_{n,\bm{k+b}}^{\uparrow}\right\rangle _{I}+{\displaystyle \sum_{\alpha}}\left\langle u_{m,\bm{k}}^{\downarrow*}|u_{n,\bm{k+b}}^{\uparrow}\right\rangle _{MT^{\alpha}}.
\end{equation}
 Within the interstitial region,

\begin{eqnarray}
\left\langle u_{m,\bm{k}}^{\downarrow*}|u_{n,\bm{k+b}}^{\uparrow}\right\rangle _{I} & = & \sum_{{\displaystyle j}}\sum_{{\displaystyle j'}}z_{m\bm{k},j}^{\downarrow}z_{n\bm{k+b},j'}^{\uparrow}\frac{1}{\Omega}\int_{cell}e^{\left[i\left(\bm{K}_{j}+\bm{K}_{j'}\right)\cdot\bm{r}\right]}\Theta\left(\bm{r}\right)d^{3}r\nonumber \\
 & = & \sum_{{\displaystyle j}}\sum_{{\displaystyle j'}}z_{m\bm{k},j}^{\downarrow}z_{n\bm{k+b},j'}^{\uparrow}\Theta\left[-\left(\bm{K}_{j}+\bm{K}_{j'}\right)\right],
\end{eqnarray}
 while inside the muffin-tin region ($\alpha$-th atom sphere),

\begin{align}
\left\langle u_{m,\bm{k}}^{\downarrow*}|u_{n,\bm{k+b}}^{\uparrow}\right\rangle _{MT^{\alpha}} & =4\pi e^{-i\left(2\bm{k+b}\right)\cdot\bm{\tau}^{\alpha}}{\displaystyle \sum_{l^{\prime\prime}m^{\prime\prime}}}\left(-i\right)^{l^{\prime\prime}}Y_{l^{\prime\prime}m^{\prime\prime}}^{*}\left(\widehat{2\bm{k+b}}\right)\nonumber \\
 & \times{\displaystyle \sum_{lm}\sum_{l^{\prime}m^{\prime}}}\left\{ A_{lm}^{\downarrow\alpha}\left(m,\bm{k}\right)A_{l^{\prime}m^{\prime}}^{\uparrow\alpha}\left(n,\bm{k+b}\right)\left[u_{l,1}^{\downarrow,\alpha}u_{l^{\prime},1}^{\uparrow,\alpha}j_{l^{\prime\prime},2k+b}\right]\right.\nonumber \\
 & +A_{lm}^{\downarrow\alpha}\left(m,\bm{k}\right)B_{l^{\prime}m^{\prime}}^{\uparrow\alpha}\left(n,\bm{k+b}\right)\left[u_{l,1}^{\downarrow,\alpha}\dot{u}_{l^{\prime},1}^{\uparrow,\alpha}j_{l^{\prime\prime},2k+b}\right]\nonumber \\
 & +A_{lm}^{\downarrow\alpha}\left(m,\bm{k}\right)C_{l^{\prime}m^{\prime}}^{\uparrow\alpha}\left(n,\bm{k+b}\right)\left[u_{l,1}^{\downarrow,\alpha}u_{l^{\prime},2}^{\uparrow,\alpha}j_{l^{\prime\prime},2k+b}\right]\nonumber \\
 & +A_{lm}^{\downarrow\alpha}\left(m,\bm{k}\right)D_{l^{\prime}m^{\prime}}^{\uparrow\alpha}\left(n,\bm{k+b}\right)\left[u_{l,1}^{\downarrow,\alpha}u_{l^{\prime},1/2}^{\uparrow,\alpha}j_{l^{\prime\prime},2k+b}\right]\nonumber \\
 & +B_{lm}^{\downarrow\alpha}\left(m,\bm{k}\right)A_{l^{\prime}m^{\prime}}^{\uparrow\alpha}\left(n,\bm{k+b}\right)\left[\dot{u}_{l,1}^{\downarrow,\alpha}u_{l^{\prime},1}^{\uparrow,\alpha}j_{l^{\prime\prime},2k+b}\right]\nonumber \\
 & +B_{lm}^{\downarrow\alpha}\left(m,\bm{k}\right)B_{l^{\prime}m^{\prime}}^{\uparrow\alpha}\left(n,\bm{k+b}\right)\left[\dot{u}_{l,1}^{\downarrow,\alpha}\dot{u}_{l^{\prime},1}^{\uparrow,\alpha}j_{l^{\prime\prime},2k+b}\right]\nonumber \\
 & +B_{lm}^{\downarrow\alpha}\left(m,\bm{k}\right)C_{l^{\prime}m^{\prime}}^{\uparrow\alpha}\left(n,\bm{k+b}\right)\left[\dot{u}_{l,1}^{\downarrow,\alpha}u_{l^{\prime},2}^{\uparrow,\alpha}j_{l^{\prime\prime},2k+b}\right]\nonumber \\
 & +B_{lm}^{\downarrow\alpha}\left(m,\bm{k}\right)D_{l^{\prime}m^{\prime}}^{\uparrow\alpha}\left(n,\bm{k+b}\right)\left[\dot{u}_{l,1}^{\downarrow,\alpha}u_{l^{\prime},1/2}^{\uparrow,\alpha}j_{l^{\prime\prime},2k+b}\right]\nonumber \\
 & +C_{lm}^{\downarrow\alpha}\left(m,\bm{k}\right)A_{l^{\prime}m^{\prime}}^{\uparrow\alpha}\left(n,\bm{k+b}\right)\left[u_{l,2}^{\downarrow,\alpha}u_{l^{\prime},1}^{\uparrow,\alpha}j_{l^{\prime\prime},2k+b}\right]\nonumber \\
 & +C_{lm}^{\downarrow\alpha}\left(m,\bm{k}\right)B_{l^{\prime}m^{\prime}}^{\uparrow\alpha}\left(n,\bm{k+b}\right)\left[u_{l,2}^{\downarrow,\alpha}\dot{u}_{l^{\prime},1}^{\uparrow,\alpha}j_{l^{\prime\prime},2k+b}\right]\nonumber \\
 & +C_{lm}^{\downarrow\alpha}\left(m,\bm{k}\right)C_{l^{\prime}m^{\prime}}^{\uparrow\alpha}\left(n,\bm{k+b}\right)\left[u_{l,2}^{\downarrow,\alpha}u_{l^{\prime},2}^{\uparrow,\alpha}j_{l^{\prime\prime},2k+b}\right]\nonumber \\
 & +C_{lm}^{\downarrow\alpha}\left(m,\bm{k}\right)D_{l^{\prime}m^{\prime}}^{\uparrow\alpha}\left(n,\bm{k+b}\right)\left[u_{l,2}^{\downarrow,\alpha}u_{l^{\prime},1/2}^{\uparrow,\alpha}j_{l^{\prime\prime},2k+b}\right]\nonumber \\
 & +D_{lm}^{\downarrow\alpha}\left(m,\bm{k}\right)A_{l^{\prime}m^{\prime}}^{\uparrow\alpha}\left(n,\bm{k+b}\right)\left[u_{l,1/2}^{\downarrow,\alpha}u_{l^{\prime},1}^{\uparrow,\alpha}j_{l^{\prime\prime},2k+b}\right]\nonumber \\
 & +D_{lm}^{\downarrow\alpha}\left(m,\bm{k}\right)B_{l^{\prime}m^{\prime}}^{\uparrow\alpha}\left(n,\bm{k+b}\right)\left[u_{l,1/2}^{\downarrow,\alpha}\dot{u}_{l^{\prime},1}^{\uparrow,\alpha}j_{l^{\prime\prime},2k+b}\right]\nonumber \\
 & +D_{lm}^{\downarrow\alpha}\left(m,\bm{k}\right)C_{l^{\prime}m^{\prime}}^{\uparrow\alpha}\left(n,\bm{k+b}\right)\left[u_{l,1/2}^{\downarrow,\alpha}u_{l^{\prime},2}^{\uparrow,\alpha}j_{l^{\prime\prime},2k+b}\right]\nonumber \\
 & \left.+D_{lm}^{\downarrow\alpha}\left(m,\bm{k}\right)D_{l^{\prime}m^{\prime}}^{\uparrow\alpha}\left(n,\bm{k+b}\right)\left[u_{l,1/2}^{\downarrow,\alpha}u_{l^{\prime},1/2}^{\uparrow,\alpha}j_{l^{\prime\prime},2k+b}\right]\right\} \left(-1\right)^{m}G_{ll^{\prime}l^{\prime\prime}}^{-mm^{\prime}m^{\prime\prime}}.
\end{align}

Finally, we consider the matrix element $\left\langle \Theta u_{m,\bm{k}}|\Theta u_{n,\bm{k+b}}\right\rangle $,

\begin{eqnarray}
\left\langle \Theta u_{m,\bm{k}}|\Theta u_{n,\bm{k+b}}\right\rangle  & = & \left\langle u_{m,\bm{k}}^{\downarrow*}|u_{n,\bm{k+b}}^{\downarrow*}\right\rangle +\left\langle u_{m,\bm{k}}^{\uparrow*}|u_{n,\bm{k+b}}^{\uparrow*}\right\rangle ,
\end{eqnarray}
 and take $\left\langle u_{m,\bm{k}}^{\downarrow*}|u_{n,\bm{k+b}}^{\downarrow*}\right\rangle $
as example, it also can be divided into two parts

\begin{equation}
\left\langle u_{m,\bm{k}}^{\downarrow*}|u_{n,\bm{k+b}}^{\downarrow*}\right\rangle =\left\langle u_{m,\bm{k}}^{\downarrow*}|u_{n,\bm{k+b}}^{\downarrow*}\right\rangle _{I}+{\displaystyle \sum_{\alpha}}\left\langle u_{m,\bm{k}}^{\downarrow*}|u_{n,\bm{k+b}}^{\downarrow*}\right\rangle _{MT^{\alpha}}.
\end{equation}
 Within the interstitial region,

\begin{eqnarray}
\left\langle u_{m,\bm{k}}^{\downarrow*}|u_{n,\bm{k+b}}^{\downarrow*}\right\rangle _{I} & = & \sum_{{\displaystyle j}}\sum_{{\displaystyle j'}}z_{m\bm{k},j}^{\downarrow}z_{n\bm{k+b},j'}^{\downarrow*}\frac{1}{\Omega}\int_{cell}e^{\left[i\left(\bm{K}_{j}-\bm{K}_{j'}\right)\cdot\bm{r}\right]}\Theta\left(\bm{r}\right)d^{3}r\nonumber \\
 & = & \sum_{{\displaystyle j}}\sum_{{\displaystyle j'}}z_{m\bm{k},j}^{\downarrow}z_{n\bm{k+b},j'}^{\downarrow*}\Theta\left[-\left(\bm{K}_{j}-\bm{K}_{j'}\right)\right],
\end{eqnarray}
 while inside the muffin-tin region ($\alpha$-th atom sphere),

\begin{align}
\left\langle u_{m,\bm{k}}^{\downarrow*}|u_{n,\bm{k+b}}^{\downarrow*}\right\rangle _{MT^{\alpha}} & =4\pi e^{i\bm{b}\cdot\bm{\tau}^{\alpha}}{\displaystyle \sum_{l^{\prime\prime}m^{\prime\prime}}}i^{l^{\prime\prime}}Y_{l^{\prime\prime}m^{\prime\prime}}^{*}\left(\widehat{\bm{b}}\right)\nonumber \\
 & \times{\displaystyle \sum_{lm}\sum_{l^{\prime}m^{\prime}}}\left\{ A_{lm}^{\downarrow\alpha}\left(m,\bm{k}\right)\left[A_{l^{\prime}m^{\prime}}^{\downarrow\alpha}\left(n,\bm{k+b}\right)\right]^{*}\left[u_{l,1}^{\downarrow,\alpha}u_{l^{\prime},1}^{\downarrow,\alpha}j_{l^{\prime\prime},b}\right]\right.\nonumber \\
 & +A_{lm}^{\downarrow\alpha}\left(m,\bm{k}\right)\left[B_{l^{\prime}m^{\prime}}^{\downarrow\alpha}\left(n,\bm{k+b}\right)\right]^{*}\left[u_{l,1}^{\downarrow,\alpha}\dot{u}_{l^{\prime},1}^{\downarrow,\alpha}j_{l^{\prime\prime},b}\right]\nonumber \\
 & +A_{lm}^{\downarrow\alpha}\left(m,\bm{k}\right)\left[C_{l^{\prime}m^{\prime}}^{\downarrow\alpha}\left(n,\bm{k+b}\right)\right]^{*}\left[u_{l,1}^{\downarrow,\alpha}u_{l^{\prime},2}^{\downarrow,\alpha}j_{l^{\prime\prime},b}\right]\nonumber \\
 & +A_{lm}^{\downarrow\alpha}\left(m,\bm{k}\right)\left[D_{l^{\prime}m^{\prime}}^{\downarrow\alpha}\left(n,\bm{k+b}\right)\right]^{*}\left[u_{l,1}^{\downarrow,\alpha}u_{l^{\prime},1/2}^{\downarrow,\alpha}j_{l^{\prime\prime},b}\right]\nonumber \\
 & +B_{lm}^{\downarrow\alpha}\left(m,\bm{k}\right)\left[A_{l^{\prime}m^{\prime}}^{\downarrow\alpha}\left(n,\bm{k+b}\right)\right]^{*}\left[\dot{u}_{l,1}^{\downarrow,\alpha}u_{l^{\prime},1}^{\downarrow,\alpha}j_{l^{\prime\prime},b}\right]\nonumber \\
 & +B_{lm}^{\downarrow\alpha}\left(m,\bm{k}\right)\left[B_{l^{\prime}m^{\prime}}^{\downarrow\alpha}\left(n,\bm{k+b}\right)\right]^{*}\left[\dot{u}_{l,1}^{\downarrow,\alpha}\dot{u}_{l^{\prime},1}^{\downarrow,\alpha}j_{l^{\prime\prime},b}\right]\nonumber \\
 & +B_{lm}^{\downarrow\alpha}\left(m,\bm{k}\right)\left[C_{l^{\prime}m^{\prime}}^{\downarrow\alpha}\left(n,\bm{k+b}\right)\right]^{*}\left[\dot{u}_{l,1}^{\downarrow,\alpha}u_{l^{\prime},2}^{\downarrow,\alpha}j_{l^{\prime\prime},b}\right]\nonumber \\
 & +B_{lm}^{\downarrow\alpha}\left(m,\bm{k}\right)\left[D_{l^{\prime}m^{\prime}}^{\downarrow\alpha}\left(n,\bm{k+b}\right)\right]^{*}\left[\dot{u}_{l,1}^{\downarrow,\alpha}u_{l^{\prime},1/2}^{\downarrow,\alpha}j_{l^{\prime\prime},b}\right]\nonumber \\
 & +C_{lm}^{\downarrow\alpha}\left(m,\bm{k}\right)\left[A_{l^{\prime}m^{\prime}}^{\downarrow\alpha}\left(n,\bm{k+b}\right)\right]^{*}\left[u_{l,2}^{\downarrow,\alpha}u_{l^{\prime},1}^{\downarrow,\alpha}j_{l^{\prime\prime},b}\right]\nonumber \\
 & +C_{lm}^{\downarrow\alpha}\left(m,\bm{k}\right)\left[B_{l^{\prime}m^{\prime}}^{\downarrow\alpha}\left(n,\bm{k+b}\right)\right]^{*}\left[u_{l,2}^{\downarrow,\alpha}\dot{u}_{l^{\prime},1}^{\downarrow,\alpha}j_{l^{\prime\prime},b}\right]\nonumber \\
 & +C_{lm}^{\downarrow\alpha}\left(m,\bm{k}\right)\left[C_{l^{\prime}m^{\prime}}^{\downarrow\alpha}\left(n,\bm{k+b}\right)\right]^{*}\left[u_{l,2}^{\downarrow,\alpha}u_{l^{\prime},2}^{\downarrow,\alpha}j_{l^{\prime\prime},b}\right]\nonumber \\
 & +C_{lm}^{\downarrow\alpha}\left(m,\bm{k}\right)\left[D_{l^{\prime}m^{\prime}}^{\downarrow\alpha}\left(n,\bm{k+b}\right)\right]^{*}\left[u_{l,2}^{\downarrow,\alpha}u_{l^{\prime},1/2}^{\downarrow,\alpha}j_{l^{\prime\prime},b}\right]\nonumber \\
 & +D_{lm}^{\downarrow\alpha}\left(m,\bm{k}\right)\left[A_{l^{\prime}m^{\prime}}^{\downarrow\alpha}\left(n,\bm{k+b}\right)\right]^{*}\left[u_{l,1/2}^{\downarrow,\alpha}u_{l^{\prime},1}^{\downarrow,\alpha}j_{l^{\prime\prime},b}\right]\nonumber \\
 & +D_{lm}^{\downarrow\alpha}\left(m,\bm{k}\right)\left[B_{l^{\prime}m^{\prime}}^{\downarrow\alpha}\left(n,\bm{k+b}\right)\right]^{*}\left[u_{l,1/2}^{\downarrow,\alpha}\dot{u}_{l^{\prime},1}^{\downarrow,\alpha}j_{l^{\prime\prime},b}\right]\nonumber \\
 & +D_{lm}^{\downarrow\alpha}\left(m,\bm{k}\right)\left[C_{l^{\prime}m^{\prime}}^{\downarrow\alpha}\left(n,\bm{k+b}\right)\right]^{*}\left[u_{l,1/2}^{\downarrow,\alpha}u_{l^{\prime},2}^{\downarrow,\alpha}j_{l^{\prime\prime},b}\right]\nonumber \\
 & \left.+D_{lm}^{\downarrow\alpha}\left(m,\bm{k}\right)\left[D_{l^{\prime}m^{\prime}}^{\downarrow\alpha}\left(n,\bm{k+b}\right)\right]^{*}\left[u_{l,1/2}^{\downarrow,\alpha}u_{l^{\prime},1/2}^{\downarrow,\alpha}j_{l^{\prime\prime},b}\right]\right\} \left(-1\right)^{m+m^{\prime}}G_{ll^{\prime}l^{\prime\prime}}^{-m-m^{\prime}m^{\prime\prime}}.
\end{align}


\begin{thebibliography}{References}
\bibitem{Kane2005a}C. L. Kane and E. J. Mele, Phys. Rev. Lett. \textbf{95},
146802 (2005).

\bibitem{Kane2005b} C. L. Kane and E. J. Mele, Phys. Rev. Lett. \textbf{95},
226801(2005).

\bibitem{Moore2010}J. E. Moore, Nature (London) \textbf{464}, 194
(2010).

\bibitem{Qi2010}X.-L. Qi and S.-C. Zhang, Physics Today \textbf{63},
33 (2010).

\bibitem{Qi2011}X.-L. Qi and S.-C. Zhang arXiv:1008.2026.

\bibitem{Hasan2010}M. Z. Hasan and C. L. Kane, Rev. Mod. Phys. \textbf{82},
3045 (2010).

\bibitem{Teo2008}J. C. Y. Teo, Liang Fu, and C. L. Kane, Phys. Rev.
B \textbf{78}, 045426 (2008).

\bibitem{Hsieh2008}D. Hsieh, D. Qian, L. Wray, Y. Xia, Y. S. Hor,
R. J. Cava, and M. Z. Hasan, Nature (London) \textbf{452}, 970 (2008).

\bibitem{Bernevig2006}B. A. Bernevig, T. L. Hughes, and S.-C. Zhang,
Science \textbf{314}, 1757 (2006).

\bibitem{Konig2007}M. K\"{o}nig, S. Wiedmann, C. Br\"{u}ne, A. Roth, H. Buhmann,
L. W. Molenkamp, X.-L. Qi, and S.-C. Zhang, Science \textbf{318},
766 (2007).

\bibitem{Fu2007a}L. Fu and C. L. Kane, Phys. Rev. B \textbf{76},
045302 (2007).

\bibitem{Zhang2009}H. Zhang, C.-X. Liu, X.-L. Qi, X. Dai, Z. Fang,
and S.-C. Zhang, Nature Phys. \textbf{5}, 438 (2009).

\bibitem{Xia2009}Y. Xia, D. Qian, D. Hsieh, L. Wray, A. Pal, H. Lin,
A. Bansil, D. Grauer, Y. S. Hor, R. J. Cava, and M. Z. Hasan, Nature
Phys. \textbf{5}, 398 (2009).

\bibitem{Chen2009}Y. L. Chen, J. G. Analytis, J.-H. Chu, Z. K. Liu,
S.-K. Mo, X. L. Qi, H. J. Zhang, D. H. Lu, X. Dai, Z. Fang, S. C.
Zhang, I. R. Fisher, Z. Hussain, and Z.-X. Shen, Science \textbf{325},
178(2009).

\bibitem{Xiao2010a}D. Xiao, Y. Yao, W. Feng, J. Wen, W. Zhu, X.-Q.
Chen, G. M. Stocks, and Z. Zhang, Phys. Rev. Lett. \textbf{105}, 096404
(2010).

\bibitem{Chadov2010}S. Chadov, X. Qi, J. K\"{u}bler, G. H. Fecher, C.
Felser, and S. C. Zhang, Nature Mater. \textbf{9}, 541 (2010).

\bibitem{Lin2010a}H. Lin, L. A. Wray, Y. Xia, S. Xu, S. Jia, R. J.
Cava, A. Bansil, and M. Z. Hasan, Nature Mater. \textbf{9}, 546 (2010).

\bibitem{Lin2010b}H. Lin, R. S. Markiewicz, L. A. Wray, L. Fu, M.
Z. Hasan, and A. Bansil, Phys. Rev. Lett. \textbf{105}, 036404 (2010).

\bibitem{Yan2010a}B. Yan, C.-X. Liu, H.-J. Zhang, C.-Y. Yam, X.-L.
Qi, T. Frauenheim, and S.-C. Zhang, Europhys. Lett. \textbf{90}, 37002
(2010).

\bibitem{Chen2010}Y. L. Chen, Z. K. Liu, J. G. Analytis, J.-H. Chu,
H. J. Zhang, B. H. Yan, S.-K. Mo, R. G. Moore, D. H. Lu, I. R. Fisher,
S. C. Zhang, Z. Hussain, and Z.-X. Shen, Phys. Rev. Lett. \textbf{105},
266401 (2010).

\bibitem{Sato2010}T. Sato, K. Segawa, H. Guo, K. Sugawara, S. Souma,
T. Takahashi, and Y. Ando, Phys. Rev. Lett. \textbf{105}, 136802 (2010).

\bibitem{Kuroda2010}K. Kuroda, M. Ye, A. Kimura, S. V. Eremeev, E.
E. Krasovskii, E. V. Chulkov, Y. Ueda, K. Miyamoto, T. Okuda, K. Shimada,
H. Namatame, and M. Taniguchi, Phys. Rev. Lett. \textbf{105}, 146801
(2010).

\bibitem{Feng2011}W. Feng, D. Xiao, J. Ding, and Y. Yao, Phys. Rev.
Lett. \textbf{106}, 016402 (2011).

\bibitem{Sun2010}Y. Sun, X.-Q. Chen, S. Yunoki, D. Li, and Y. Li,
Phys. Rev. Lett. \textbf{105}, 216406 (2010).

\bibitem{Yan2010b}B. Yan, H.-J. Zhang, C.-X. Liu, X.-L. Qi, T. Frauenheim,
and S.-C. Zhang, Phys. Rev. B \textbf{82}, 161108 (2010).

\bibitem{Kim2010}J. Kim, J. Kim, and S.-H. Jhi, Phys. Rev. B \textbf{82},
201312 (2010).

\bibitem{Jin2011}H. Jin, J.-H. Song, A. J. Freeman, M. G. Kanatzidis,
Phys. Rev. B \textbf{83}, 041202 (2011).

\bibitem{Zhang2011}H.-J. Zhang, S. Chadov, L. M\"{u}chler, B. Yan, X.-L.
Qi, J. K\"{u}bler, S.-C. Zhang, and C. Felser, Phys. Rev. Lett. \textbf{106},
156402 (2011).

\bibitem{Chen2011}S. Chen, X. G. Gong, C.-G. Duan, Z.-Q. Zhu, J.-H.
Chu, A. Walsh, Y.-G. Yao, J. Ma, and S.-H. Wei, Phys. Rev. B \textbf{83},
245202 (2011).

\bibitem{Fu2007b}L. Fu, C. L. Kane and E. J. Mele, Phys. Rev. Lett.
\textbf{98}, 106803 (2007).

\bibitem{Moore2007}J. E. Moore and L. Balents, Phys. Rev. B \textbf{75},
121306 (2007).

\bibitem{Roy2009}R. Roy, Phys. Rev. B \textbf{79}, 195322 (2009).

\bibitem{Fu2006}L. Fu and C. L. Kane, Phys. Rev. B \textbf{74}, 195312
(2006).

\bibitem{Fukui2007}T. Fukui and Y. Hatsugai, J. Phys. Soc. Jpn. \textbf{76},
053702 (2007).

\bibitem{Xiao2010b}D. Xiao, M.-C. Chang, and Q. Niu, Rev. Mod. Phys.
\textbf{82}, 1959 (2010).

\bibitem{Liu2011}C.-C. Liu, W. Feng, and Y. Yao, arXiv:1104.1290,
Phys. Rev. Lett. (in press).

\bibitem{Soluyanov2011}A. A. Soluyanov and D. Vanderbilt, Phys. Rev.
B \textbf{83}, 235401 (2011).

\bibitem{Yu2011}R. Yu, X.-L. Qi, A. Bernevig, Z. Fang, and X. Dai,
arXiv:1101.2011.

\bibitem{Singh1994}D. J. Singh, \textit{Planewaves, Pseudopotentials
and the LAPW Method} (Kluwer Academic, Boston, 1994).

\bibitem{Blugel2006}S. Bl\"{u}gel and G. Bihlmayer, John von Neumann
Institute for Computing, NIC Series \textbf{31}, 85 (2006).

\bibitem{Blaha2001}P. Blaha, K. Schwarz, G. Madsen, D. Kvaniscka,
and J. Luitz, \textit{Wien2k, An Augmented Plane Wave Plus Local Orbitals
Program for Calculating Crystal Properties} (Vienna University of
Technology, Vienna, Austria, 2001).

\bibitem{Junes2001}J. Kune\v{s}, P. Nov\'{a}k, R. Schmid, P. Blaha, and K.
Schwarz, Phys. Rev. B \textbf{64}, 153102 (2001).

\bibitem{King-Smith1993}R. D. King-Smith and D. Vanderbilt, Phys.
Rev. B \textbf{47}, 1651 (1993).

\bibitem{Resta1994}R. Resta, Rev. Mod. Phys. \textbf{66}, 899 (1994).

\bibitem{Perdew1996}J. P. Perdew, K. Burke, and M. Ernzerhof, Phys.
Rev. Lett. \textbf{77}, 3865 (1996).

\bibitem{Tran2009}F. Tran and P. Blaha, Phys. Rev. Lett. \textbf{102},
226401 (2009).

\bibitem{Feng2010}W. Feng, D. Xiao, Y. Zhang, and Y. Yao, Phys. Rev.
B \textbf{82}, 235121 (2010).

\bibitem{Al-Sawai2010}W. Al-Sawai, H. Lin, R. S. Markiewicz, L. A.
Wray, Y. Xia, S.-Y. Xu, M. Z. Hasan, and A. Bansil, Phys. Rev. B \textbf{82},
125208 (2010).

\bibitem{Haase2002}M. G. Haase, T. Schmidt, C. G. Richter, H. Block
and W. Jeitschko, J. Solid State Chem. \textbf{168}, 18 (2002).

\bibitem{Pamplin1979}B. R. Pamplin, T. Kiyosawa, and K. Masumoto,
Prog. Cryst. Growth Charact. \textbf{1}, 331 (1979).

\bibitem{Limpijumnong2002}S. Limpijumnong, and W. R. L. Lambrecht,
Phys. Rev. B \textbf{65}, 165204 (2002). \end{thebibliography}
\end{document}